\documentclass{cplslarge}

\usepackage{deluxetable} 
\usepackage{amsmath}
\usepackage{amssymb}

\usepackage{mediabb}

\usepackage[authoryear]{natbib}
\usepackage{afterpage}
\usepackage{aas_macros}

\begin{document}
\editedlistofcontributors
\frontmatter
\mainmatter
\setcounter{page}{1} 
   \author[Kasuga \&\  Jewitt]{Toshihiro Kasuga and David Jewitt}
   \setcounter{chapter}{7}
   \chapter{\bf Asteroid--Meteoroid Complexes}

\section{Introduction}
Physical disintegration of asteroids and comets leads to the production of orbit-hugging debris streams.  In many cases, the mechanisms underlying disintegration are uncharacterized, or even unknown.  Therefore, considerable scientific interest lies in tracing the physical and dynamical properties of the asteroid-meteoroid complexes 
backwards in time, in order to learn how they form.

Small solar system bodies offer the opportunity to understand the origin and evolution of the planetary system.   They include the comets and asteroids, as well as the mostly unseen objects in the much more distant Kuiper belt and Oort cloud reservoirs.
Observationally, asteroids and comets are distinguished principally by their optical morphologies, with asteroids appearing as point sources, 
and comets as diffuse objects with unbound atmospheres (comae), at least when near the Sun.  
The principal difference between the two is thought to be the volatile content, especially the abundance of water ice.  
Sublimation of ice in comets drives a gas flux into the adjacent vacuum while drag forces from the expanding gas are exerted on embedded dust and debris particles, expelling them into interplanetary space.  Meteoroid streams, consisting of large particles ejected at low speeds and confined to move approximately in the orbit of the parent body, are one result.  

\index{ablation}
When the orbit of the parent body intersects that of the Earth, meteoroids strike the atmosphere 
and all but the largest are burned up by frictional heating, creating the familiar meteors \citep{O34}.  
Later, the phenomenon has been realized as ablation by shock wave radiation heating \citep{Bron83}.
The first established comet-meteoroid stream relationships were  
identified by G. Schiaparelli and E. Weiss in 1866 \citep[see][]{CBE98}. 
In the last twenty years, a cometary meteoroid stream theory has been established enabling accurate  
shower activity prediction of both major \citep[e.g.~Leonids,][]{KMR97,MA99JI} and 
minor showers \citep[2004 June ${\rm Bo\ddot{o}tids}$,][]{VAS05}.  
This theory deals with the perturbed motion of streams encountering 
the Earth after the ejection from relevant parent bodies.
This improved theory has  provided striking opportunities for  
meteor shower studies of orbital trajectories, velocities and compositions, 
resulting in a revolution in meteor science.     

\index{IAU Meteor Data Center}
\index{complex!asteroid-meteoroid complex}
\index{meteoroid stream}
Some meteoroid streams seem to be made of debris released from 
asteroids.  The notion that not all  stream parents are comets is comparatively old,   
having been suggested by \cite{W39me,W39Gem,W40} \citep[see also][]{O25,H37}.
The Geminid meteoroid stream \citep[GEM/4, from][]{JopekJenniskens11}$\footnote{IAU Meteor Data Center, 
Nomenclature}$ and asteroid 3200 Phaethon 
are probably the best-known examples \citep{W83}.   In such cases, it appears unlikely that
ice sublimation drives the expulsion of solid matter, raising the general question of what produces the meteoroid streams? 
Suggested alternative triggers include thermal stress, rotational instability and collisions (impacts) by secondary bodies \citep[]{J12,JHA15}.  
Any of the above, if sufficiently violent or prolonged, could lead to the production of a debris trail that would, if it crossed Earth's orbit, be classified as a meteoroid stream or an ``Asteroid-Meteoroid Complex", 
comprising streams and several macroscopic, split fragments \citep{VK86, Jones86, CBE98}.

\index{D-criterion} 
\index{Lidov-Kozai mechanism} 
The dynamics of stream members and their parent objects may differ, and dynamical associations are not always obvious.  Direct searches for dynamical similarities employ a distance parameter, $D_{\rm SH}$, which measures the separation in  orbital element space by comparing $q$ (perihelion distance), 
$e$ (eccentricity), $i$ (inclination), $\Omega$ (longitude of the ascending node), and 
$\omega$ (argument of perihelion) \citep[]{SW63}.
A smaller $D_{\rm SH}$ indicates a closer degree of orbital similarity between two bodies, 
with an empirical cut-off for significance often set at $D_{\rm SH} \lesssim$ 0.10--0.20 \citep[][Section 9.2.2]{chapter9}. 
The statistical significance of proposed parent-shower associations 
has been coupled with $D_{\rm SH}$ \citep{WiegertPB04,Ye2016}.
Recent models assess the long-term dynamical stability for high-$i$ and -$e$ asteroids.  
\cite{OSK06,OAI08} find the Phaethon-Geminid Complex (PGC) 
and the Icarus complex together using as criteria the $C_1$ \citep{M45} and $C_2$ \citep{Lidov62} integrals.  
These are secular orbital variations expressed by 

\begin{eqnarray}
C_1 & = & (1 - e^2)\,{\rm cos}^2(i) \label{C1} \\
C_2 & = & e^2\,(0.4 - {\rm sin}^2(i)\,{\rm sin}^2(\omega)).
\label{C2}
\end{eqnarray}

So-called time-lag theory is utilized to demonstrate long-term orbital evolution of complex members.  
When a stream-complex is formed, 
the orbital energies ($\propto$\,$a^{-1}$, where $a$ is the semimajor axis) of ejected fragments are expected to be slightly different from the energy of the precursor.  
The motions of the released objects are either  
accelerated or decelerated relative to the precursor under gravitational perturbations 
(possibly including non-gravitational perturbations), effectively
causing a time lag, $\Delta$$t$, in the orbital evolution to arise. 
Both $C_1$ and $C_2$ are approximately invariant during dynamical evolution, 
distinguishing the complex members.  
The PGC members (Phaethon, 2005 UD, 1999 YC), for example, dynamically follow
the Lidov-Kozai mechanism based on secularly perturbed motion of the asteroids \citep{Kozai62}.

\index{near-Earth objects (NEOs)}
\index{Jupiter family comets (JFCs)}
\index{Halley type comets (HTCs)}
\index{Encke-type comets}
\index{meteoroid stream!Geminids}
\index{meteoroid stream!Quadrantids}
\index{complex!Taurid Complex}
\index{meteoroid stream!Phoenicids}
\index{meteoroid stream!Leonids}
\index{meteoroid stream!Perseids}
\index{Kozai resonance}
Most known parent bodies are near-Earth Objects (NEOs), including both asteroids and comets.  
The comets include various sub-types: Jupiter family comets (JFCs), Halley type comets (HTCs) and Encke-type comets \citep[][]{Ye2018} (Table~\ref{orbit}).  
A classification of parents and their associated streams has been proposed based on their inferred evolutionary stages \citep{Babadzhanov_Obrubov87,Babadzhanov_Obrubov91,Babadzhanov_Obrubov92CeMDA,Babadzhanov_Obrubov92}. 
Some streams originating from asteroids (e.g. Phaethon), or from  comets (e.g. 96P, 2P) are the most evolved.   
For example, the Geminids, Quadrantids (QUA/10) and Taurid Complex (TAU/247) show stable secular variation of the orbital elements under the mean motion resonance 
with Jupiter and the Kozai resonance, producing annual meteor showers.   
On the other hand, young streams are usually from JFCs and HTCs.  
JFCs orbits, in particular, are chaotically scattered by frequent close encounters with Jupiter, 
tending to produce irregular streams, e.g. Phoenicids (PHO/254).  
HTCs orbit with widespread inclinations, including retrograde orbits that are absent in the JFCs, 
and may also generate regular showers, e.g. Leonids (LEO/13) and Perseids (PER/7).

Non-gravitational force effects can be important in the evolution of stream complexes but are difficult to model, since they depend on many unknowns such as the size, rotation, and thermal properties of the small bodies involved.  In the last decade, video- and radar- based surveys of meteors have also been used to trace the trajectories back to 
potential parent NEOs, and so to find new stream complexes
\citep[e.g.][]{BWW08II,BWW08,BWWW10,Musci2012,Jenniskens08,WerykBrown12,Rudawska2015,Jenniskens2016a,Jenniskens2016c,Jenniskens2016b,Ye2016} \citep[Reviewed in][]{Jenniskens17}.

\index{spectroscopy}
\index{abundance!solar}
\index{abundance!elemental}
\index{intensity ratio}
\index{sodium (Na)} 
Meteor spectroscopy provides some additional constraints on the composition of meteoroids  \citep{MM63, Mill80}.   
Spectroscopy is typically capable of obtaining useful data for meteors having   optical absolute magnitudes
 +3 to +4 or brighter, corresponding to meteoroid 
sizes $\gtrsim$ 1 mm \citep[][]{Lindblad1987,CBE98,BKS10}.  
Meteor spectra consist primarily of atomic emission lines and 
some molecular bands in the visible to near-infrared wavelengths.  
The commonly identified neutral atoms are Mg\,{\sc i}, Fe\,{\sc i}, Ca\,{\sc i}, and Na\,{\sc i}, while 
the singly ionized atomic emissions of Ca\,{\sc ii} and Mg\,{\sc ii} also appear in some fast-moving meteors (e.g.~Leonids with geocentric velocity $V_g \sim$72 km\,${\rm s^{-1}}$ impact much more quickly than the Geminids, with $V_g \sim$ 35\,km\,s$^{-1}$).  
The abundances, the excitation temperatures and the electron densities can be deduced for each spectrum,  assuming the Boltzmann distribution for electron energies, but the measurements are difficult, and the resulting elemental abundances and/or intensity ratios \cite[e.g.][]{Naga78,Bo93,KYW05,BKS05} 
are  scattered \citep[summarized in][]{CBE98,KYKW06}.  
Generally, most meteors are found to have solar abundance within factors of $\sim$3--4. 
Some  elements  are noticeably underabundant, probably affected by 
incomplete evaporation \citep{TLBF03,BSB99,KYW05}.  In particular, sodium (Na) is a relatively abundant and moderately volatile element that is easily volatilized.
As a result, the abundance of Na in meteors is a good indicator of thermal evolution of meteoroids. 
Heating either during their residence in interplanetary space or within the parent bodies themselves can lead to a sodium depletion.  

\index{meteoroid stream}
In this chapter, we give a brief summary of observational results on parents and their associated showers.
We discuss the properties of specific complexes, and tabulate their dynamical and physical properties (Tables~\ref{orbit}, \ref{physical}). 
We focus on the main meteoroid streams for which the properties and associations seem the most secure.  Numerous additional streams and their less certain associations are discussed in the literature reviewed by \cite[][Section 7.5]{PJ06,Jenniskens2008EMP,Boro07,Ye2018,chapter7}.

   
\section{General Properties}

\begin{figure*}
\figurebox{32pc}{}{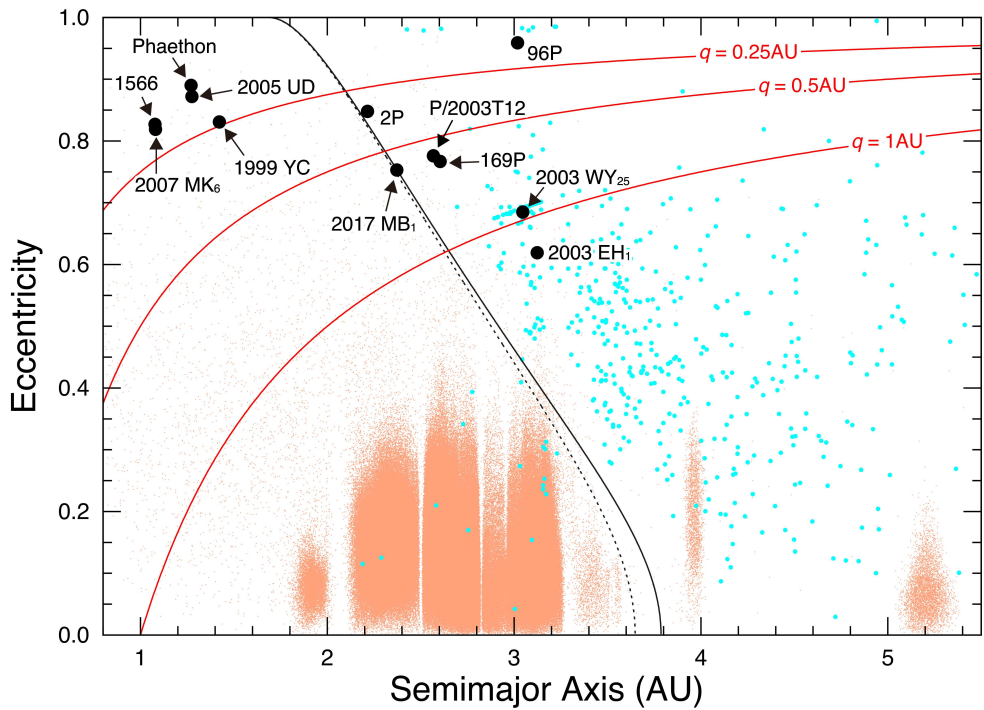}   
\caption{Distribution of the parent bodies (black circles) 
in the semimajor axis vs. orbital eccentricity plane (cf. Table~\ref{orbit}).   
The distributions of asteroids (brown dots) and 
comets (light-blue dots) are shown for reference.   
The lines for $T_{\rm J}$ = 3.08 with $i$=0$^{\circ}$ (solid curve) and with $i$=9$^{\circ}$ (dotted curve) 
broadly separate asteroids and comets.  
Perihelion distances $q$ = 0.25, 0.5 and 1\,AU are shown as red curves.    
\label{ae}}
\vspace*{7pc} 
\end{figure*}

\begin{figure*}
\figurebox{32pc}{}{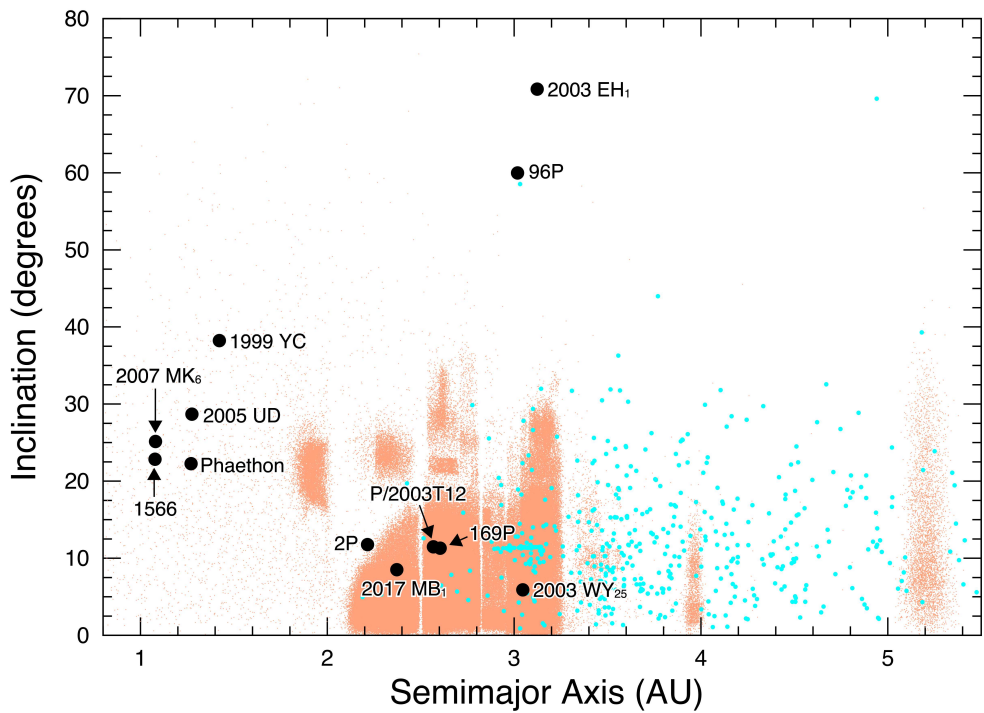}   
\caption{Same as Figure \ref{ae} but semimajor axis vs. inclination.
\label{ai}}
\vspace*{2pc} 
\end{figure*}

\index{complex!asteroid-meteoroid complex} 
To date, twelve objects in the six asteroid-meteoroid complexes have been studied in detail. 
Figure \ref{ae} represents their distributions in the semimajor axis versus eccentricity plane,  
while Figure \ref{ai} shows semimajor axis versus inclination 
(Table~\ref{orbit} summarizes the orbital properties).

\begin{table*}
\caption{Orbital Properties \label{orbit}}{\tabcolsep5.5pt%
\vspace*{-2ex}
\begin{tabular}{@{}llcccccccccccc@{}}
\toprule
Complex & Object        &  $a$\tablenotemark{a} &  $e\tablenotemark{b}$    & $i$\tablenotemark{c}    &   $q$\tablenotemark{d}   & $\omega$\tablenotemark{e}  & $\Omega$\tablenotemark{f}  &  $Q$\tablenotemark{g}  & $P_{\rm orb}\tablenotemark{h}$ & $C_1$\tablenotemark{i}   & $C_2 $\tablenotemark{i} &  $T_{\rm J}$\tablenotemark{j}\\
\hline
Geminids    & Phaethon        & 1.271 & 0.890 & 22.253 & 0.140 & 322.174 & 265.231 & 2.402 & 1.43 & 0.178 & 0.274&  4.509 \\
            & 2005 UD         & 1.275 & 0.872 & 28.682 & 0.163 & 207.573 & 19.746  & 2.387 & 1.44 & 0.184 & 0.267 & 4.504 \\	
            & 1999 YC         & 1.422 & 0.831 & 38.226 & 0.241 & 156.395 & 64.791  & 2.603 & 1.70 & 0.191 & 0.234 & 4.114 \\
Quadrantids & 2003 EH$_1$     & 3.123 & 0.619 & 70.838 & 1.191 & 171.361 & 282.979 & 5.056 & 5.52 & 0.066 & 0.146 & 2.065 \\
            & 96P/Machholz 1  & 3.018 & 0.959 & 59.975 & 0.125 & 14.622  & 94.548  & 5.911 & 5.24 & 0.020 & 0.324 & 1.939 \\
Capricornids& 169P/NEAT       & 2.604 & 0.767 & 11.304 & 0.607 & 217.977 & 176.219 & 4.602 & 4.20 & 0.396 & 0.227 & 2.887 \\    
            & P/2003 T12      & 2.568 & 0.776 & 11.475 & 0.575 & 217.669 & 176.465 & 4.561 & 4.12 & 0.382 & 0.232 & 2.894 \\	 
            & 2017 MB$_1$     & 2.372 & 0.753 &  8.508 & 0.586 & 264.628 & 126.974 & 4.158 & 3.65 & 0.424 & 0.215 & 3.071 \\ 
Taurids     & 2P/Encke        & 2.215 & 0.848 & 11.781 & 0.336 & 186.542 & 334.569 & 4.094 & 3.30 & 0.269 & 0.287 & 3.025 \\    
Taurids-Perseids & 1566 Icarus& 1.078 & 0.827 & 22.852 & 0.187 & 31.297  & 88.082  & 1.969 & 1.12 & 0.268 & 0.246 & 5.296 \\
                 & 2007 MK$_6$& 1.081 & 0.819 & 25.138 & 0.196 & 25.466  & 92.887  & 1.966 & 1.12 & 0.270 & 0.246 & 5.284 \\    
Phoenicids  & 2003 WY$_{25}$  & 3.046 & 0.685 & 5.9000 & 0.961 & 9.839   &  68.931 & 5.132 & 5.32 & 0.525 & 0.188 & 2.816 \\
\hline    
\end{tabular}}
\begin{tabnote}
Notes: Orbital data are obtained from NASA JPL HORIZONS (https://ssd.jpl.nasa.gov/horizons.cgi)\\
 $^a$ Semimajor axis (AU)\\
 $^b$ Eccentricity \\
 $^c$ Inclination (degrees)\\
 $^d$ Perihelion distance (AU)\\
 $^e$ Argument of perihelion (degrees)\\
 $^f$ Longitude of ascending node (degrees)\\
 $^g$ Aphelion distance (AU)\\
 $^h$ Orbital period (yr)\\
 $^i$ Dynamical invariants: $C_1$ = (1 -- $e^2$)\,${\rm cos}^2(i)$, $C_2$ = $e^2$\,(0.4 -- ${\rm sin}^2(i)$\,${\rm sin}^2$($\omega$))\\
 $^j$ Tisserand parameter with respect to Jupiter (asteroids with $T_J > 3.08$, comets with $T_J < $ 3.08; $a < a_J$(5.2\,AU))\\
\end{tabnote}
\end{table*}

\begin{table*}
\caption{Physical Properties \label{physical}}{\tabcolsep5.5pt%
\begin{tabular}{@{}llcccccccc@{}}
\toprule%
Complex & Object   &  $D_e$\tablenotemark{a}  &  $p_v$\tablenotemark{b}   & $P_{\rm rot}$\tablenotemark{c}    &   $B-V$\tablenotemark{d}   & $V-R$\tablenotemark{d}  & $R-I$\tablenotemark{d}  
            &  $a/b$\tablenotemark{e}  \\
\hline
Geminids & Phaethon$^1$ & 5--6        & 0.09--0.13 & 3.604 & 0.59$\pm$0.01 & 0.35$\pm$0.01 & 0.32$\pm$0.01 & $\sim$1.45 \\
         & 2005 UD$^2$  & 1.3$\pm$0.1 & 0.11$^{f}$    & 5.249 & 0.66$\pm$0.03 & 0.35$\pm$0.02 & 0.33$\pm$0.02 & 1.45$\pm$0.06 \\	
         & 1999 YC$^3$  & 1.7$\pm$0.2$^{g}$ & 0.09$\pm$0.03$^{g}$ & 4.495 & 0.71$\pm$0.04 & 0.36$\pm$0.03 &       -       & 1.89$\pm$0.09 \\
Quadrantids & 2003 EH$_1$$^4$ & 4.0$\pm$0.3 & 0.04$^{f}$    &12.650 & 0.69$\pm$0.01 & 0.39$\pm$0.01 & 0.38$\pm$0.01 & 1.50$\pm$0.01 \\
            & 96P/Machholz 1$^5$ & 6.4   & 0.04$^{f}$   & 6.38 &      -        & 0.40$\pm$0.03 &       -       & 1.4\,$\le$ \\
Capricornids& 169P/NEAT$^6$ & 4.6$\pm$0.6 & 0.03$\pm$0.01    & 8.410  & 0.73$\pm$0.02& 0.43$\pm$0.02 & 0.44$\pm$0.04 & 1.31$\pm$0.03 \\    
            & P/2003 T12 & -         &  -            &  -     &   -   &      -               &       -       &  -  \\	 
            & 2017 MB$_1$$^7$& 0.52  &  -            &  6.69  &   -   &      -               &       -       &  $\sim$1.2  \\ 
Taurids     & 2P/Encke$^8$  & 4.8$\pm$0.4      & 0.05$\pm$0.02&   11  & 0.73$\pm$0.06 & 0.39$\pm$0.06 &       -           & 1.44$\pm$0.06 \\    
Taurids-Perseids & 1566 Icarus$^9$& 1.0--1.3     & 0.30--0.50$^{h}$ & 2.273 & 0.76$\pm$0.02 & 0.41$\pm$0.02 &  0.28$\pm$0.02    & 1.2--1.4\\
                 & 2007 MK$_6$   &  0.18$^{i}$        & 0.40$^{f}$  &   -   &      -        &     -         &       -       &  -       \\    
Phoenicids  & 2003 WY$_{25}$$^{10}$  & $\le$\,0.32     & 0.04$^{f}$&   -   &      -        &     -         &       -       &  -  \\
\hline    
\end{tabular}}
\begin{tabnote}
Notes: \\
$^{a}$ Effective diameter (km)\\
$^{b}$ Geometric albedo\\
$^{c}$ Rotational period (hr)\\
$^{d}$ Color index. Solar colors are $B-V$=0.64$\pm$0.02, $V-R$=0.35$\pm$0.01 and $R-I$=0.33$\pm$0.01 \citep{HFP06}.\\
$^{e}$ Axis ratio\\
$^{f}$ Assumed value\\
$^{g}$ $D_e$=1.4$\pm$0.1 with the assumed $p_v$=0.11 \citep{KJ08}. \\
$^{h}$ Extremely high $p_v$ $>$ 0.7 is given $D_e <$ 0.8\,km \citep{Mahapatra99,HarrisLagerros02}.\\
$^{i}$ Estimated from the absolute magnitude $H$=20.3 (MPO386777) with the assumed $p_v$=0.40.\\
$^1$ \cite{Green85,Tedesco04,Dundon05,AMH14,Hanus16,Taylor18}
$^2$ \cite{JH06,KOS07}
$^3$ \cite{KJ08,Mainzer11,War17}
$^4$ \cite{KJ15}
$^5$ \cite{LTL00,MHM04,LTF04}
$^6$ \cite{KBW10,WAL04,AH05,DeMeoBinzel08}
$^7$ \cite{Warner18MPBu}
$^8$ \cite{KSL17,LTF04,CF02,LOW07,FLU00}
$^9$ \cite{GRT70, VHM89,Chapman94,NMM15,MY69,Harris98,DeAn95,Dundon05} 
$^{10}$ \cite{J06}
\end{tabnote}
\end{table*}

\index{Tisserand parameter}
\index{3-body problem}
\index{Kuiper belt comets}
\index{Oort cloud comets}
Traditionally, the Tisserand parameter with respect to Jupiter, $T_J$, 
is used to characterize the dynamics of small bodies \citep{Kresak82,Kozai92}.  It is defined by

\begin{equation}
T_J = \frac{a_J}{a} + 2\left[(1-e^2)\frac{a}{a_J}\right]^{1/2}\cos(i)
\label{tisserand}
\end{equation}

\noindent where $a$, $e$ and $i$ are the semimajor axis, eccentricity and inclination of the orbit and $a_J$ = 5.2 AU is the semimajor axis of the orbit of Jupiter.  This parameter, which is conserved in the circular, restricted 3-body problem, provides a measure of the close-approach speed to Jupiter. Jupiter itself has $T_J$ = 3. Main belt asteroids have $a < a_J$ and  $T_J >$ 3 while dynamical comets from the Kuiper belt have 2 $\le T_J <$ 3 and comets from the Oort cloud have $T_J < 2$.   In principle, the asteroids and comets can also be distinguished compositionally. The main belt asteroids are generally rocky, non-icy objects which probably formed inside snow-line in the protoplanetary disk, while comets contain a larger ice fraction  and formed beyond it.   In practice, it is difficult or impossible to measure the compositions of most small bodies in the solar system, so that composition is not often a useful diagnostic.

The use of $T_J$ as a discriminant breaks down near $T_J$ $\simeq$ 3,  since  the definition assumes that Jupiter's orbit is a circle, the gravity of other planets is neglected, and so on.    Accordingly, a functional definition for the boundary employed here is 
$T_J$ $\simeq$ 3.08 \citep{JHA15}, shown  in Figure \ref{ae}, where the solid curve represents  $T_J$ computed assuming $i$ = 0$^{\circ}$, while the 
dotted curve is the same but with  $i$ = 9$^{\circ}$ (equal to the average inclination of 516,633 numbered objects in the JPL Horizons database).    
They are very similar, indicating the definition broadly works independently of $i$.
This avoids chaotic cases caused by deviation of the real solar system from the circular, 
restricted 3-body approximation.  Also, comet 2P/Encke with $T_J$ $\sim$ 3.03 and 
the quasi-Hilda comets with $T_J$ $\sim$ 2.9 -- 3.04 are appropriately classified with this criterion.

\index{near-Sun objects} 
As seen in Figure~\ref{ae}, all of the parent bodies have $e \gtrsim$ 0.6.   
Seven objects fall on the right side of the boundary with $T_J$ $<$ 3.08, 
corresponding to the region of comets.
Mass-loss activity has been directly detected in most of them, although 2003~EH$_1$ and 2017~MB$_1$ have yet to show evidence for current activity.  
Objects on the left side of the diagram are classified in the region of the asteroids, with T$_J$ $>$ 3.08.  
In these objects,  a range of physical processes appear  to drive the mass loss.   
They have small $q \lesssim$ 0.25\,AU and are categorized as near-Sun objects (see Figure~\ref{ae}).  
Recurrent activity of Phaethon at perihelion, including the formation of a tail, has been reported.  

In Figure \ref{ai},  2003~EH$_1$ and 96P show remarkably high-$i$ $\gtrsim$ 60$^{\circ}$.  
Five objects with $a$ = 1.0 $\sim$ 1.4\,AU have moderate $i$ of 20$^\circ$ $\sim$ 40$^\circ$.  
Another five objects at $a$ = 2.2 $\sim$3.1\,AU have low-$i$, 
compatible with those of most main belt asteroids.

\section{Known Asteroid-Meteoroid Complexes}
\label{complexes}

The physical properties of the main asteroid-meteoroid complexes are listed in Table~\ref{physical}, 
and we review them focusing on the observational evidence.

\subsection{Geminids -- (3200) Phaethon}
\label{PGC}

\begin{figure}
\figurebox{13.8pc}{}{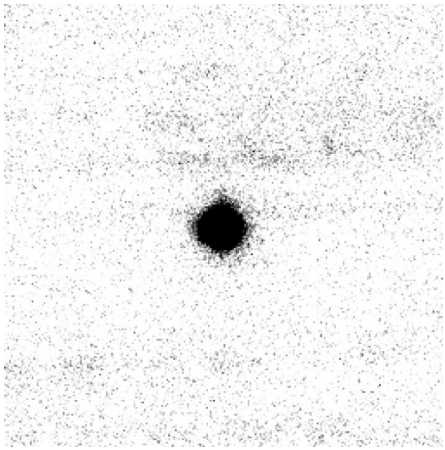}
\caption{Composite R-band image of Phaethon taken at the University of Hawai'i 2.2-m telescope (UH2.2) on UT 2003 December 19. 
The full image width is 1'  and this is a 3600\,s equivalent integration, with Phaethon at heliocentric distance $R$ = 1.60\,AU, 
geocentric distance $\Delta$ = 1.39\,AU, and 
phase angle $\alpha$ = 0.3$^{\circ}$.
The image shows no evidence of mass-loss activity. 
From \cite{HJ05}. \label{Phaethon}}
\end{figure}

\index{meteoroid stream!Geminids}
\index{asteroid!(3200)~Phaethon} 
The Geminid meteor shower is one of the most active annual showers \citep[][]{W39Gem,Sup93}. 
The shower currently has zenithal hourly rate (ZHR), 
the number of meteors visible per hour under a clear-dark sky ($\lesssim$ limiting magnitude +6.5), 
of $\sim$ 120 and is expected to continue to increase  to a peak ZHR $\sim$ 190 in 2050 as the Earth moves deeper into 
the stream core \citep{JH86,PJ06}.
The Geminids are dynamically associated with the near-Earth asteroid (3200) Phaethon (1983\,TB) \citep{W83} (Figure~\ref{Phaethon}). 
A notable orbital feature of both is 
the small perihelion distance, $q\sim$ 0.14\,AU, raising the possibility of strong thermal processing of the surface.   
Indeed, the peak temperature at perihelion is $\sim$1000\,K \citep{ONN09}.

%
\index{spectroscopy}
\index{sodium (Na)}
\index{intensity ratio}
\index{abundance!solar} 
Spectroscopy of Geminid meteors shows an extreme diversity in Na content,   
from strong depletion of Na abundance in some to sun-like values in others.  
\cite{KWE05} found a huge depletion in the Na/Mg abundance ratio of a Geminid meteor, 
with a value an order of magnitude smaller than the solar abundance ratio \citep{AG89,L03}.  
Line intensity ratios of Na\,{\sc i}, Mg\,{\sc i}, and Fe\,{\sc i} emissions
of the Geminids also show a wide  range of Na line strengths, from undetectable to intense \citep{BKS05}.
Studies of Geminid meteor spectra, reported since 
the 1950s \citep[e.g.][]{M55,RSSW56,Harvey73,B01,TLBF03,TLF04}, 
mostly fit this pattern.

As a summary for Na in the Geminids in the last decade, \cite{KYKW06} 
investigated perihelion dependent thermal effects on meteoroid streams.  
The effect is supposed to alter the metal abundances from their intrinsic values in their parents, 
especially for temperature-sensitive elements: a good example is Na in alkaline silicate.  
As a result, meteoroid streams with $q \lesssim 0.1$\,AU should be 
depleted in Na by thermal desorption,   
because the corresponding meteoroid temperature (characterized as blackbody)  
exceeds the sublimation temperature of alkaline silicates ($\sim$ 900\,K for sodalite: ${\rm Na_{4}(AlSiO_{4})_{3}Cl}$).   
For this reason, the Na loss in Geminids ($q\sim$ 0.14\,AU) is 
most likely to be caused by thermal processes on Phaethon itself (see section~\ref{Na}).  

\begin{figure*}
\figurebox{36pc}{}{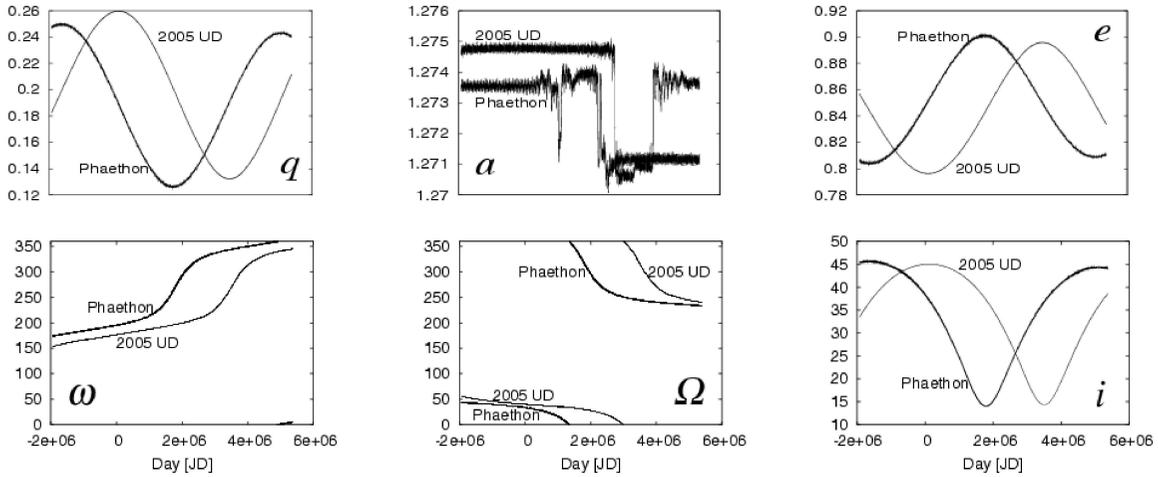}
\caption{Dynamical evolution of (3200)~Phaethon and 2005~UD 
for the six orbital elements, representing similar behavior with a time-shift of $\sim$4600 yr. 
The abscissa shows time in Julian Terrestrial Date (JTD).
From \cite{OSK06}
\label{DyUD}}
\vspace*{4pc}
\end{figure*}

\index{Trojan!Jovian}
The parent body Phaethon \citep[diameter 5 -- 6\,km, from][]{Tedesco04,Taylor18}
has an optically blue (so-called ``B-type'') 
reflection spectrum that distinguishes it from most other asteroids and from the nuclei of comets.   
Specifically, only $\sim$ 1 in 23 asteroids is of B-spectral type and most cometary nuclei are slightly reddish, 
like the C-type asteroids or the (redder) D-type Jovian Trojans.

\begin{figure}
\figurebox{17pc}{}{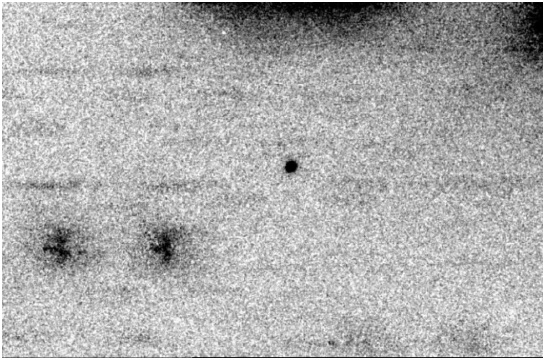}
\caption{R-band image of 2005~UD in 1800\,s exposure recorded using the UH2.2 on UT 2005 November 21. 
The region shown is $\sim$ 150" in width and the distances and phase angle of the object were $R$ = 1.59\,AU,  $\Delta$ = 0.96\,AU and $\alpha$ = 35.8$^{\circ}$, respectively.  
From \cite{JH06}.
\label{UD}}
\end{figure}

\begin{figure}
\figurebox{17pc}{}{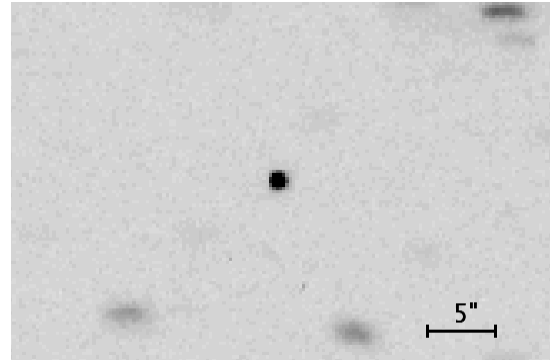}
\caption{Keck telescope R-band image of 1999~YC on UT 2007 October 12 showing a point source 
with $FWHM \sim$ 0.65" in 400\,s integration, centered within a frame 40'' wide.
Heliocentric, geocentric distances and phase angle were $R$ = 2.60\,AU,  $\Delta$ = 1.82\,AU 
and $\alpha$ = 16.4$^{\circ}$, respectively.
From \cite{KJ08}.
\label{YC}}
\end{figure}

\index{taxonomy!B-type}
\index{asteroid!(2)~Pallas}
\index{asteroid!(3200)~Phaethon}
\index{asteroid!2005~UD}
\index{asteroid!1999~YC}
\index{color}
\index{complex!Phaethon-Geminid Complex (PGC)} 
A dynamical pathway to another B-type, the large main-belt  asteroid (2)~Pallas, 
has been reported with $\sim$ 2\% probability (\citeauthor{deLeon10}, \citeyear{deLeon10}, see also \citeauthor{NT18}, \citeyear{NT18}).
However, the colors of Phaethon and Pallas are not strictly identical, 
as would be expected if one were an unprocessed chip from the other, although both are blue.  
Color  differences might result from preferential heating and modification of Phaethon, 
with its much smaller perihelion distance (0.14 AU vs. $\sim$2.1 AU for Pallas).  
Recently, \cite{OSK06,OAI08} suggested the existence of  
a ``Phaethon-Geminid  Complex (PGC)''--consisting of
a group of dynamically associated split fragments, and identified 
the 1\,km-sized asteroids 2005~UD and 1999~YC as 
having a common origin with Phaethon (cf. Figure~\ref{DyUD}).  
Photometry of 2005~UD and 1999~YC revealed the optical colors (Figures~\ref{UD}, \ref{YC}).  
The former is another rare blue object \citep{JH06, KOS07} while the latter is spectrally neutral \citep{KJ08}
(see section \ref{colors}).

%
\index{meteoroid stream!Geminids}
The key question is how the Geminid meteoroid stream was produced from Phaethon.  
The extreme possibilities are that the Geminids are the products of a catastrophic event (for example an energetic collision, or a rotational disruption) or that they are produced in steady-state by continuing mass-loss from Phaethon.

\index{Jupiter family comets (JFCs)} 
In the steady-state case, the entire stream mass, $M_s \sim 10^{12}$ -- $10^{13}$\,kg \citep{HM89,PJ94,Bla17}\citep[cf. 1--2 orders larger in][]{Ryabova17}, must be released over the last $\tau \sim 10^3$ years (the dynamical lifetime of the stream, cf.~\cite[][Section 7.5.4]{J78,JH86,Gu89,WW93,Ryabova99,Beech02,JakubNeslu15,chapter7}). 
This gives $dM_s/dt \sim M_s/\tau$ = 30 -- 300 kg s$^{-1}$, comparable to the mass loss rates exhibited by active Jupiter family comets.   However, while the Jupiter family comets are notable for their distinctive comae of ejected dust, Phaethon generally appears as a point source, devoid of coma or other evidence for on-going mass loss \citep[][]{CMSS96,HJ05,WHR08} (cf. Figure~\ref{Phaethon}).

\begin{figure}
\figurebox{20pc}{}{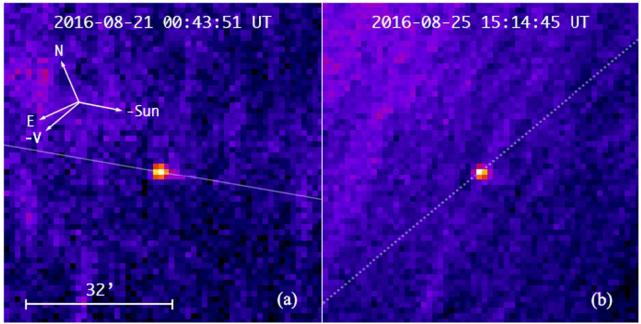}
\caption{
STEREO-A image taken on 2016 August 20  
of (a) (3200)~Phaethon and tail aligned on the position angles of the anti-solar direction
and (b) the whole image sequence on the negative heliocentric velocity vector projected on the sky plane. 
Geometric parameters are $R$ $\sim$ 0.14\,AU, Phaethon-STEREO distance $\sim$ 0.9\,AU, and the angle $\sim$ 120$^{\circ}$.  
The projected Sun-Phaethon line (the narrow white line across the left panel) is drawn to 
better illustrate that the direction of the tail was anti-solar, 
while the white dotted line across the right panel is the orientation. 
From \cite{HL17}.
\label{Pha16}}
\end{figure}

\index{spacecraft!Solar Terrestrial Relations Observatory (STEREO)} 
Recently, this general picture has changed with the detection of mass loss using near-perihelion observations  taken
with the Solar Terrestrial Relations Observatory (STEREO) spacecraft in 2009, 2012 and 2016 
\citep{JL10,JLA13,LJ13,HL17}.  
In addition to factor-of-two brightening at perihelion relative to the expected phase-darkened inverse-square law brightness, a diffuse, linear tail has been resolved, as shown in Figure~\ref{Pha16}.
Because of its close association with the high temperatures experienced at perihelion, the activity is likely to result from thermal fracture or the desiccation of hydrated minerals. The observed particles from Phaethon are micron-sized, and are highly susceptible to solar radiation pressure sweeping.  They are rapidly accelerated by radiation pressure and so cannot be retained in the Geminid stream \citep{JL10}.   In addition, their combined mass, $\sim$ 3 $\times$ $10^5$\,kg per perihelion, is at least 10$^7 \times$ smaller than the stream mass, $M_s$ \citep{JLA13}.   It is possible that much more mass is contained in larger particles which, however, present a small fraction of the scattering cross-section and which, therefore, are unsampled in the optical data from STEREO.

\index{spacecraft!Deep Impact}
\index{spacecraft!DESTINY+} 
\index{lunar impact} 
In this regard, larger particles (sizes $>$10 $\mu$m) were recently reported 
in thermal emission at 25 $\micron$\, \citep{Arendt14}, 
but evidently contribute little to the optical scattering cross-section.  
The particles in the stream are estimated to be near mm-scale or larger \citep[e.g.][]{Bla17}, up to $\sim$1 to 10\,cm,  
as measured in lunar impacts \citep{Yanagisawa08,Suggs14,Ortiz15,Szalay18}.   
The limit to optical depth of a Phaethon's trail is $\leq$ 3$\times$10$^{-9}$ \citep[][]{Jewitt18HST}, 
consistent with those of cometary dust trails \citep[10$^{-9}$--10$^{-8}$, from][]{SykesWalker92,Ishiguro09Ad} 
(see also section~\ref{MS}).  
While continued long wavelength observations of Phaethon to detect 
large particles will be helpful \citep{JHA15}, 
the true nature of Phaethon and the PGC complex objects may await 
spacecraft missions resembling NASA's ``Deep Impact'' 
\citep{AH05,KWS06,Kasuga07Ad} and JAXA's ``DESTINY+" \citep{Sarli18}.


\subsection{Quadrantids -- 2003 EH$_1$}
\label{QUAD}

\index{meteoroid stream!Quadrantids}
\index{complex!Quadrantids} 
The Quadrantid meteor shower was first reported in 1835 \citep{Que1839} and appears annually in early January.   
The shower consists of two different components, the so-called young and old meteoroid streams, 
which represent  a very short duration of core activity (lasting $\sim$0.5 day)
and a broader, longer-lived ($\sim$4 days) background activity  \citep[][and references therein]{WB05, BWWW10}. 
The width of a meteoroid stream depends on its age, as a result of broadening by  accumulated planetary perturbations. 
The small width of the Quadrantid core stream indicates ejection ages of only $\sim$200--500 years \citep{PJ04,WRB04,ASW15}, 
and there is even some suggestion that the first reports of meteoroid stream activity coincide with the formation of the stream.  
On the other hand, the broader  background stream implies larger ages of perhaps $\sim$3,500 years or more 
\citep{OYW95, OYW08, KN07}.

\begin{figure}
\figurebox{20pc}{}{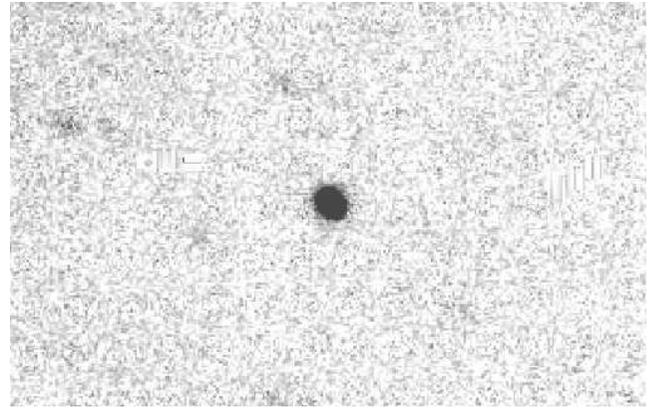}
\caption{2003~EH$_1$ in a 360\,s R-band image taken at Keck 
on UT 2013 October 2. 
No coma or tail is visible on the object having an FWHM of 0".86 
in the frame of 40'' $\times$ 25''. 
$R$=2.1\,AU, $\Delta$=2.0\,AU and $\alpha$=27.6$^{\circ}$.  
From \cite{KJ15}.
\label{EH1}}
\end{figure}

%
\index{asteroid!2003~EH$_1$}
Two parent bodies of the Quadrantid complex have been proposed.  
The 4\,km diameter Near-Earth Object (196256) 2003~EH$_1$ (hereafter 2003 EH$_1$),   
 discovered on UT 2003 March 6 by the Lowell Observatory Near-Earth-Object Search (LONEOS) \citep{Skiff03}, may be responsible for the young core stream \citep{PJ04,WRB04,WB05,baba08c,Jo11,ASW15}.
The orbit of 2003 EH$_1$ has ${\it a}$ = 3.123\,AU, ${\it e}$ = 0.619, ${\it i}$ = $70^{\circ}.838$ 
and ${\it q}$ = 1.191\,AU (Table~\ref{orbit}).  
The ${\it T_{\rm J}}$ (= 2.07) identifies it as a likely  Jupiter family comet, albeit one in which  on-going activity has yet to be detected
\citep{KBS06,baba08c,BKS09,Tanc14}.   
The steady-state production rates $\lesssim 10^{-2}$ kg ${\rm s^{-1}}$ estimated 
from 2003~EH$_1$ at $R$ = 2.1\,AU 
are at least five orders of magnitude too small to supply the core Quadrantid stream mass
$M_s \sim 10^{13}$ \,kg \citep{KJ15} (see Figure~\ref{EH1}).    
Even at $q$=0.7--0.9\,AU a few hundred years ago, sublimation-driven activity from 
the entire body takes $\sim$ 10s of years in the whole orbit, being hard to reconcile.  
In order to form the core Quadrantid stream, we consider  episodic replenishment by an unknown process to be more likely.

\index{comet!96P/Machholz~1} 
Comet 96P/Machholz~1 has been suggested as the source of the older, broader
part of the Quadrantid complex \citep[][Section 7.5.2]{chapter7}, with meteoroids released 
 2,000--5,000 years ago~\citep{Mc90,Babadzhanov_Obrubov91,GRF92,JJ93,WB05,AWP17,AWJ18}.  
Comet 96P currently has a small perihelion orbit 
($\it a$ = 3.018\,AU, $\it e$ = 0.959,  $\it i$ = $59^{\circ}.975$ and 
$\it q$ = 0.125\,AU from Table~\ref{orbit}) substantially different from that of 2003~EH$_1$.
Despite this, calculations show  rapid dynamical evolution  that allows the possibility that 2003~EH$_1$ 
is a fragment of 96P, or that both were released from a precursor 
body~\citep[together defining the Machholz complex:][]{SC05}.  
One or both of these bodies  can be the parents of 
the Quadrantid meteoroids~\citep[][Section 7.5.2]{KN07,baba08c,NHJ13,NKT13,NKT14,chapter7}.

A notable dynamical feature of 2003~EH$_1$ is the strong evolution of the perihelion distance \citep{WB05,NKT13,FG14}.
Numerical integrations indicate that the minimum perihelion distance $q$ $\sim$ 0.12\,AU (${\it e}$ $\sim$ 0.96) 
occurred $\sim$1500\,yr ago \citep[][]{NKT13,FG14}, and 
the perihelion has increased approximately linearly with time from 0.2\,AU 1000 years ago 
to the present-day value of 1.2\,AU.  
At its recently very small (Phaethon-like) perihelion distance, it is reasonable to expect that the surface layers should have been 
heated to the point of fracture and desiccation (see section~\ref{LK}).

As described above, the Phaethon-produced Geminid meteoroids ($q \sim$0.14\,AU) show extreme 
diversity in their Na abundance, from strong depletion to near sun-like Na content \citep{KWE05,BKS05}.  
Curiously, the Quadrantid meteoroids from the core stream are less depleted in Na than 
the majority of Geminid meteoroids~\citep[][]{KBS06,BKS09}.  
The interpretation of this observation is unclear (see section~\ref{AvsI}).  

The optical colors of 2003~EH$_1$ are similar to, 
but slightly redder than, those of the Sun. They are most taxonomically compatible 
with the colors of C-type asteroids \citep{KJ15} (see section \ref{colors}).

\subsection{Capricornids -- 169P/NEAT}

\index{meteoroid stream!Capricornids}
\index{complex!Capricornids} 
The $\alpha$-Capricornids (CAP/1) are active from late July to early August, usually showing slow  ($\sim$22\,km s$^{-1}$) and bright meteors.  
The shower, with an ascending nodal intersection of $\omega$=270$^{\circ}$ with the Earth, is expected to be a twin stream also  producing a daytime shower \citep{PJ06}.
Because of the low entry velocities,  the meteor plasma excitation temperature is T$_{ex} \lesssim$ 3600K and no trace of high temperature gas (i.e. hot component of T$_{ex}$ $\sim$ 10$^4$\,K) is found \citep{BW96}.  
The metal contents of the $\alpha$-Capricornids are unremarkable, being within a factor of a few of the Solar abundance \citep{BW96, MTO14}. 

\begin{figure}
\figurebox{20pc}{}{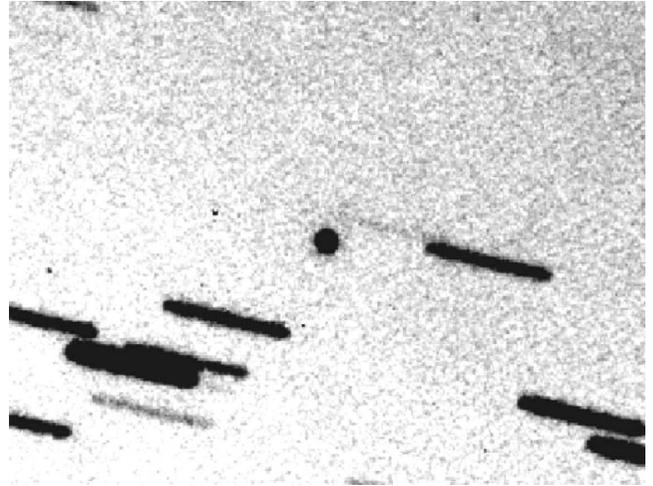}
\caption{
Comet 169P/NEAT in a 600\,sec, r'-band image
taken at the Dominion Astrophysical Observatory 1.8 m telescope on UT 2010 February 17. 
The frame size is 200$^{^{\prime \prime }}$ × 150$^{^{\prime \prime }}$. 
No coma or tail is visible on the object having an FWHM of 2.8". 
$R$=1.43\,AU, $\Delta$=0.47\,AU and $\alpha$=16.1$^{\circ}$. 
From \cite{KBW10}.
\label{169P}}
\end{figure}

\index{meteoroid stream!Daytime Capricornids-Sagittariids} 
\index{asteroid!2002~EX$_{12}$}
\index{comet!169P/NEAT} 
Recently \cite{BWWW10} suggested that the Daytime Capricornids-Sagittariids (DCS/115) are  closely dynamically related to 
the $\alpha$-Capricornids.  
One of the parent body candidates, comet 169P/NEAT, has been identified as the parent body of the $\alpha$-Capricornid meteoroid stream 
by numerical simulations ~\citep{JV10}.  
The object was discovered as asteroid 2002~EX$_{12}$ by the NEAT survey in 2002 \citep[cf.][]{WF05,Green05} and 
 was re-designated as 169P/NEAT in 2005 after revealing a cometary appearance \citep{Green05}.    
The orbital properties ~\citep[][$T_J$ = 2.89]{L96} and optical observations 
reveal that 169P/NEAT is a $\sim$ 4 km diameter, nearly dormant Jupiter family comet 
with tiny mass-loss rate $\sim$ 10$^{-2}$\,kg\,s$^{-1}$ \citep{KBW10} (Figure~\ref{169P}). 

In the steady state, the stream mass $M_s$ $\sim$10$^{13}$--10$^{15}$\,kg and the age $\tau$$\sim$5,000\,yr \citep{JV10} together
require a mass-loss rate four orders of magnitude larger than  measured in 2010.      
In the case of 169P, cometary activity (a dust tail) was confirmed in 2005 
(almost 1 orbital period before) and episodic mass-loss should be expected.  
This is a different case from other asteroidal parents of complexes. 
\cite{KBW10} used the fractional change in the spin angular velocity to 
estimate the mass loss from 169P as $\sim$ 10$^{9}$--10$^{10}$\,kg per orbit.  
With the $M_s$ and $\tau$, the conclusion is that 
the origin of the $\alpha$-Capricornids meteoroid 
stream could be formed by the steady disintegration of 169P.

\index{comet!P/2003~T12 (SOHO)}
\index{asteroid!2017~MB$_1$} 
Other parent body candidates continue to be proposed for the Capricornids.  
P/2003~T12 (SOHO) was suggested to share a common parent with 169P, following a breakup $\sim$2900\,yr ago \citep{SF15}.
Comet 169P is a large, almost inactive body \citep{KBW10}, while 
P/2003~T12 seems to be a very small comet, with a sub-km radius nucleus \citep{SF15} 
accompanied by dust-tails in near-Sun STEREO-B observations \citep{HMT13}.   
The orbit of 2017~MB$_1$ was suggested to resemble that of the $\alpha$-Capricornids meteor shower \citep{WCCB17}.
2017~MB$_1$ has not been reported to show any sign of mass-loss activity.

\subsection{Taurid Complex -- 2P/Encke}
\label{2P}

\index{meteoroid stream!Taurids}
\index{complex!Taurid Complex}
\index{comet!2P/Encke} 
The Taurid meteor shower includes the Northern, the Southern and other small branches \citep[][Section 7.5.5]{chapter7}, 
possibly originating from more than one parent body.   
The Taurids show protracted,  low-level activity with many fireballs 
from September to December,  peaking in early November every year.     
The Taurid meteoroid complex has been suggested to be formed by a 
disrupted giant comet (40\,km-sized) 10$^4$ years ago \citep{CN84,CN87,ACS93}, although the very recent break-up of such a large (i.e.~rare) body is statistically unlikely.  Comet 2P/Encke has, for a long time, been considered as the most probable parent of the shower \citep{W40}.

The Taurid complex has a dispersed structure with low inclination and perihelia between 0.2 and 0.5 \,AU. 
The low inclination of the stream orbit  enhances the effect of planetary perturbations 
from the terrestrial planets \citep{LTW06}, 
resulting in the observed, diffuse structure of the complex \citep{MTR17}.  
Furthermore, 2P has a relatively small heliocentric distance (aphelion is $Q$ = 4.1AU) allowing it to 
stay mildly active around its orbit, and producing a  larger spread in $q \sim$0.34\,AU, 
than could be explained from  ejection at perihelion alone \citep{GRWK06,KSL17}.

An unfortunate artifact of the low inclination of the Taurid complex ($i \sim$ 12$^\circ$, Table~\ref{orbit}) is that many near-Earth asteroids are plausible parent bodies 
based on orbital dynamical calculations 
\citep[e.g.][]{ACS93,SA96,baba01,PWK04,PKW06,BWK08}.  
As a result, many of the proposed associations are likely coincidental \citep{SBM17,MTR17}.  
Actually no spectroscopic linkage between 2P/Encke and the 10 potential Taurid-complex NEOs 
has been confirmed (the latter are classified variously as X, S, Q, C, V, O, and K-types) \citep{Popescu14,TSM15}, this
being totally different from the case of the Phaethon-Geminid Complex.
Here, we focus on the physical properties of the Taurid meteor shower and the most strongly associated parent, 2P/Encke.

Spectroscopic studies of some Taurid meteors find a carbonaceous feature \citep{Boro07,MTR17}.   
The heterogeneity (large dispersion of Fe content) and low strength (0.02--0.10 MPa) of the Taurids \citep[][Section 2.3.4]{chapter2}
suggest a cometary origin, consistent with but not proving  2P as a parent  \citep{Boro07,MTR17}. 
Note that the current perihelion distance ($q \sim$0.34\,AU, where $T \sim$ 480 K) is too large for strong thermal metamorphism to be expected \citep{Boro07}.

\begin{figure}
\figurebox{13pc}{}{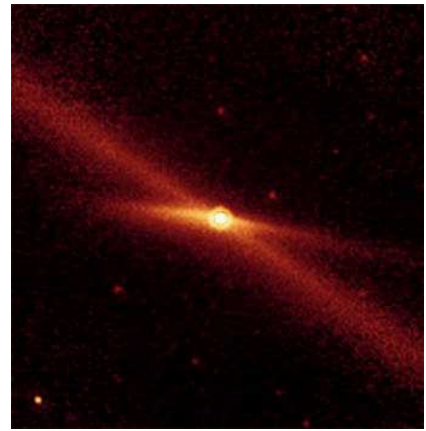}
\caption{
The 24 $\mu$m infrared image of comet 2P/Encke obtained in 2004 June by 
the Spitzer Space Telescope.  
The image field of view is about 6' and centered on the nucleus.
The near horizontal emission is produced by recent cometary activity, 
and the diagonal emission across the image is the meteoroid stream 
along the orbit (\citeauthor{Kelley06},\citeyear{Kelley06}; see also Reach et al., 2007).  
Courtesy NASA/JPL-Caltech/M. Kelley (Univ. of Minnesota).
\label{2P}}
\end{figure}

%
\index{color} 
Comet 2P/Encke is one of the best characterized short-period comets, with published determinations of its
rotation period, color, albedo and phase function \citep[see reviews;][]{LTF04,KSL17}.
For example, the effective radius is 2.4\,km, the average color indices, 
$B-V$ = 0.73$\pm$0.06 and $V-R$ = 0.39$\pm$0.06 \citep[e.g.][]{LOW07}, 
the rotational period is about 11\,hr or 22\,hr (but changing with time in response to outgassing torques \citep[][and references therein]{BSF05,KSL17}; with minimum axis ratio 
1.4 (\citeauthor{FLU00}, \citeyear{FLU00}; \citeauthor{LW07}, \citeyear{LW07}; see also, \citeauthor{LTF04}, \citeyear{LTF04}).

\index{Jupiter family comets (JFCs)}
\index{potentially hazardous meteoroid stream} 
Optically, 2P appears dust-poor because optically bright micron and sub-micron 
sized dust particles are under-abundant in its coma \citep{2PJ04}.
However, the dust~/~gas ratio determined from thermal emission is an extraordinary $\mu \sim$ 30, suggesting a dust-rich body \citep{RSL2000, LFA04} 
compared to, for example, Jupiter family comet 67P/Churyumov-Gerasimenko, where $\mu$ = 4$\pm$2  \citep[][]{RSD15}.
The total mass loss is 2--6 $\times$ 10$^{10}$\,kg per orbit, mostly in the form of large particles 
that spread around the orbit and give rise to 2P's thermal dust trail as the source of Taurid meteor showers \citep{AC97, RSL2000} (see Figure~\ref{2P}).
The entire structured stream mass, $M_s \sim 10^{14}$\,kg \citep{Asher_CN94}, must have been 
released over the last $\tau$ $\sim$ 5,000 -- 20,000\,years~\citep[the dynamical lifetime of the stream, cf.][]{W40,Babadzhanov_Obrubov92CeMDA,PJ06}.  
This gives $dM_s/dt \sim M_s/\tau$ = 200 -- 600\,kg\,s$^{-1}$, a few times larger than 
the mass loss rates typically reported for active Jupiter family comets.  
Various small near-Earth objects and some meteorite falls have been linked with the orbit of the stream as 
potentially hazardous \citep{BMM13,Olech17,SBM17}.

\subsection{Sekanina's (1973) Taurids-Perseids -- Icarus} 
\label{IcarusMK6}

\begin{figure}
\figurebox{20pc}{}{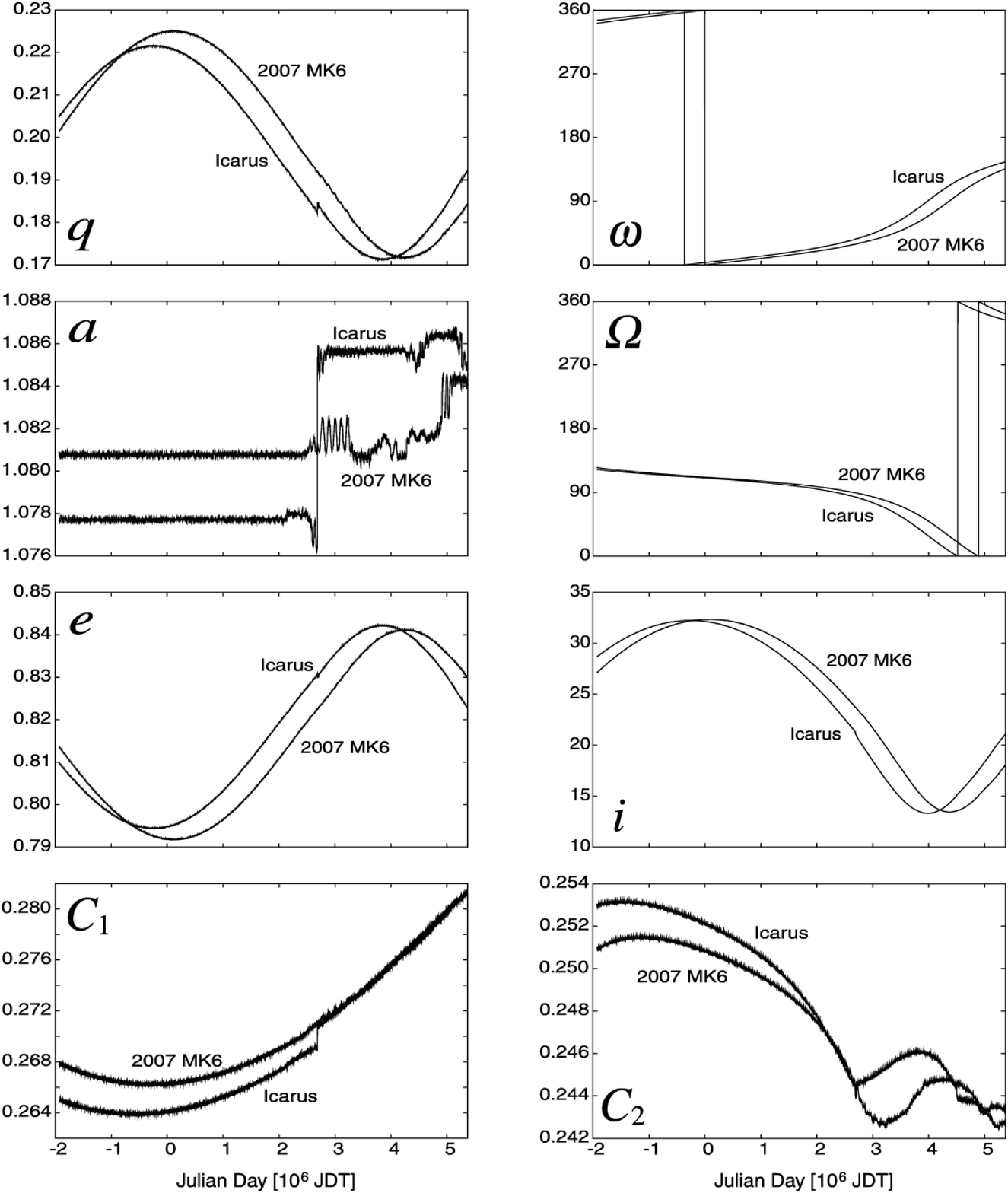}
\caption{Dynamical evolution process of 1566~Icarus and 2007~MK$_6$ in JTD. 
The orbital elements ($q$, $a$, $e$, $\omega$, $\Omega$ and $i$) 
and the $C_1$ and $C_2$ integrals are plotted (cf. Table~\ref{orbit}).  
Time-shifting is only $\sim$1,000 yr.    
From \cite{OAI07}.
\label{MK6}}
\end{figure}

\index{meteoroid stream!Sekanina's (1973) Taurid-Perseids}
\index{complex!Sekanina's (1973) Taurid-Perseids}
\index{asteroid!1566~Icarus} 
\index{asteroid!2007~MK$_6$} 
The Icarus asteroid family was reported as the first family found in the 
near-Earth region, which dynamically relates asteroids 1566~Icarus, 2007~MK$_6$ 
and Sekanina's (1973) Taurid-Perseid meteor shower \citep{OAI07} (see Figure~\ref{MK6}).

\index{asteroid!1949~MA}
Near-Earth Apollo asteroid 1566~Icarus (= 1949~MA) was discovered in 1949, 
having distinctive small $q$=0.19\,AU and high $i$=23$^{\circ}$ \citep{BCF50}.
The object has diameter $D_e$$\sim$ 1\,km \citep[e.g. Table~1, from][]{Chapman94}, a moderately high albedo of 0.30 -- 0.50 \citep[cf.][]{GRT70, VHM89, Chapman94,NMM15} 
and a short rotational period, $\sim$2.273\,hr \citep[e.g.][]{MY69} \citep[see also,][]{HarrisLagerros02}. 
A reflection spectrum close to Q- or V-type asteroids is found \citep[][]{GRT70,Hicks98,Dundon05} 
(see section \ref{colors}).

\index{asteroid!2007~MK$_6$}
\index{asteroid!2006~KT$_{67}$}
\index{meteoroid swarm} 
On the other hand, near-Earth asteroid 2007~MK$_6$ (= 2006~KT$_{67}$) 
was discovered in 2007 \citep{Hill07}. 
Assuming an albedo like that of Icarus, then  
2007~MK${_6}$ is $\sim$180\,m in diameter 
as computed from the absolute magnitude $H$=20.3 (MPO386777) (see Table~\ref{physical}).
The breakup hypothesis from Icarus, if true, could be due to near a critical rotation period 
and thermal stress induced at small $q$$\sim$0.19\,AU (subsolar temperature $\sim$ 900\,K), which might be related to 
the production of the meteoroid stream (see section \ref{Rotation}). 
The Taurid-Perseid meteoroids can be dynamically related with 
the Icarus asteroid family (D$_{\rm SH}$ $\sim$ 0.08), 
speculated to cross the Earth's orbit \citep{Seka73}.  
The rare detection of the Taurid-Perseid meteor shower may result from the intermittent stream (swarm) 
due to very limited dust supply phase from the parent body.

\section{Possible Complexes}

Here, we describe two examples of less well-characterized  complexes suspected to 
include stream branches and one or more parent bodies.

\subsection{Phoenicids -- Comet D/1819~W1 (Blanpain)}
\label{PhoWY25}

\begin{figure}
\figurebox{20pc}{}{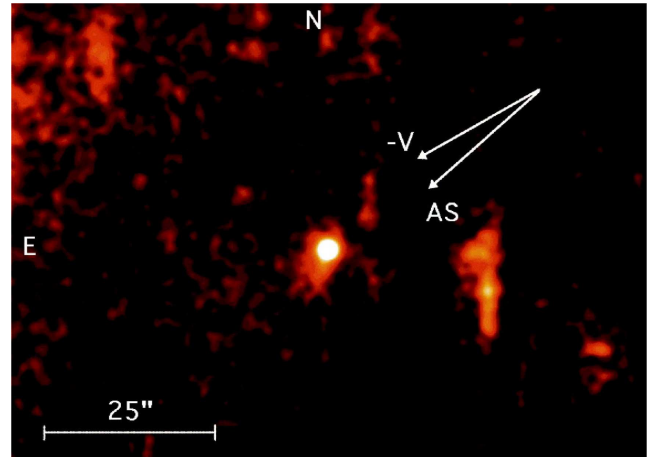}
\caption{2003~WY$_{25}$ (=289P/Blanpain) imaged in R-band, 500\,sec integration 
at $R$=1.6\,AU, $\Delta$=0.7\,AU and $\alpha$=20.7$^{\circ}$ 
using UH2.2 on UT 2004 March 20.
A faint coma is apparent, extending to the southeast. 
Arrows show the directions of the negative heliocentric velocity vector (marked ''-V") 
and the anti-solar direction (''AS").  
The estimated radius $\sim$ 160\,m is the smallest active cometary parent ever observed.  
From \cite{J06}.  
\label{WY25}}
\end{figure}

\index{meteoroid stream!Phoenicids}
\index{complex!Phoenicids}
\index{asteroid!2003~WY$_{25}$} 
\index{comet!D/1819~W1 (289P/Blanpain)}
\index{Jupiter family comets (JFCs)} 
\index{break-up}
The Phoenicid meteor shower was first reported more than 50 years ago, 
on December 5 in 1956 \citep{HN57}. 
Promptly, the lost Jupiter family comet D/1819~W1 (289P/Blanpain) was proposed as the potential 
source (\citeauthor{Ridley57}, \citeyear{Ridley57} reviewed in \citeauthor{Ridley63}, \citeyear{Ridley63}).   
In 2003, the planet-crossing asteroid 2003~WY$_{25}$ was discovered \citep{Ticha03}, 
with orbital elements resembling those of D/Blanpain \citep{FMR05,M05}.
The related Phoenicids' activity in 1956 and 2014 \citep{WSK05,PJL05,SWT17,TSW17}, 
raised the possibility that 2003~WY$_{25}$ might be either the dead nucleus of D/Blanpain 
itself or a remnant of the nucleus surviving from an earlier, unseen disintegration. 
\cite{J06} optically observed asteroid 2003~WY$_{25}$, finding the radius of 160\,m (an order of magnitude smaller than 
typical cometary nuclei), and revealing a 
 weak coma consistent with mass-loss rates of 10$^{-2}$\,kg\,s$^{-1}$ (Figure~\ref{WY25}).
The latter is too small to supply the estimated 10$^{11}$ kg stream mass on reasonable timescales \citep[$\le$10,000\,yrs,][]{PJL05}. Indeed, the mass of 2003~WY$_{25}$ (assuming density 1000\,kg\,m$^{-3}$ and a spherical shape) is $\sim$2$\times$10$^{10}$ kg, smaller than the stream mass. 
Either the stream was produced impulsively by the final stages of the break-up of a once much larger precursor to 2003~WY$_{25}$, or another parent body may await discovery \citep{J06}.

\subsection{Andromedids -- Comet 3D/Biela}

\index{meteoroid stream!Andromedids}
\index{complex!Andromedids}
\index{comet!3D/Biela}
\index{Jupiter family comets (JFCs)} 
The Andromedid meteor shower (AND/18) was firstly reported in 1798 \citep{HSS59}.  
The dynamics were linked with Jupiter family comet 3D/Biela \citep[][]{K88,K99}.  
The shower is proposed to  result from continuous disintegration  
of 3D/Biela from 1842 until its sudden disappearance in 1852, 
resulting in irregular meteor shower appearances \citep[e.g.][]{O25,Cook73,JV07}.   
The estimated stream mass is 10$^{10}$\,kg \citep{PJL05}, however, 
the absence of  parent candidates means that little can be determined about 
the production of the meteoroids. 
Nonetheless,  the Andromedid meteor shower was actually detected by radar in 2011 and 
is numerically predicted to appear in the coming decades \citep[][]{WBW13}.

   \section{Parent Bodies}

In this section we discuss group physical properties of the parent bodies (cf. Table~\ref{physical}).  
Most objects (e.g. 3200~Phaethon, 2005~UD, 1999~YC, 2003~EH$_1$ and 169P)
show point-like images (Figures~\ref{Phaethon}, \ref{UD}, \ref{YC}, \ref{EH1}, \ref{169P}) from which we can be confident that the measured properties  
refer to the bare objects (or nuclei) alone.      
However, 2P/Encke, 2003~WY$_{25}$ and some other comets may sometimes be active (e.g. Figures~\ref{2P}, \ref{WY25}) leading to potential confusion between the properties of the nucleus and the near-nucleus comae.

\index{color}
\subsection{Colors}
\label{colors}

\begin{figure}
\figurebox{20pc}{}{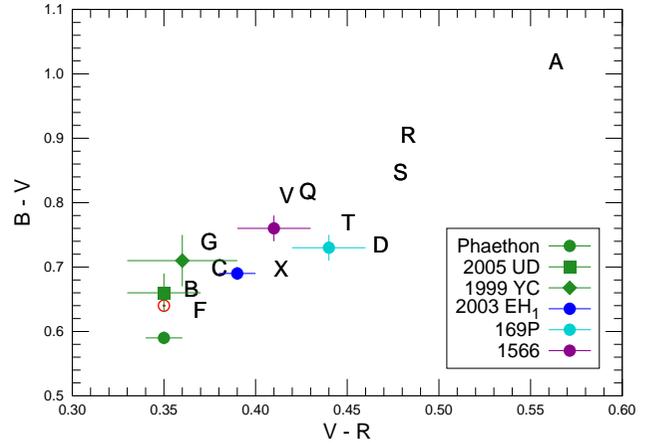}
\caption{
Color plots of $V - R$ vs. $B - V$ showing parent bodies (filled symbols) from Table~\ref{physical}, 
and Tholen taxonomic classifications \citep{Tholen84PhD}, 
as tabulated by \cite{DFC03}.  The color of the Sun (red circle) is also plotted.
\label{BVR}}
\end{figure}

\begin{figure}
\figurebox{20pc}{}{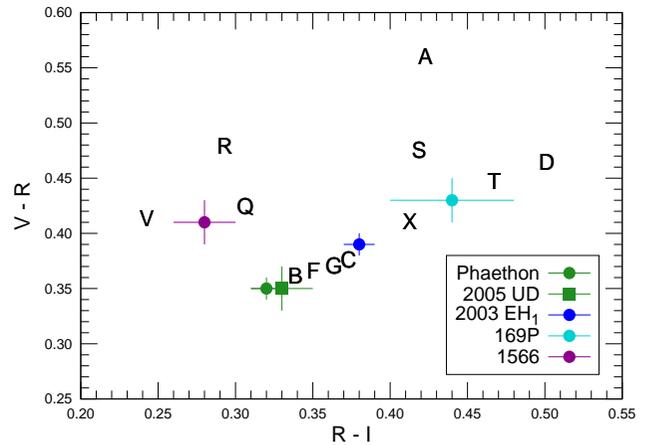}
\caption{
The same as Figure~\ref{BVR} but in the $R - I$ vs. $V - R$ color plane.  
The color of the Sun is exactly coincident with that of 2005~UD.
\label{VRI}}
\end{figure}

\index{color}
\index{Tholen taxonomy} 
Figures \ref{BVR} and \ref{VRI} show distributions of the colors of the parent bodies from Table~\ref{physical}. 
In addition, Tholen taxonomy classes are plotted from photometry of NEOs from \cite{DFC03}.
Here, 2P is not included because of the coma contamination suggesting mild activity during the whole orbit.

\index{complex!Phaethon-Geminid Complex (PGC)}
\index{asteroid!2005~UD}
\index{asteroid!1999~YC}
\index{taxonomy!B-type}
\index{taxonomy!C-type} 
The asteroids of the PGC (3200 and 2005~UD, 1999~YC) show colors from nearly neutral  to blue.  
Asteroids 3200 Phaethon and 2005~UD are classified as B-type asteroids 
\citep[cf.][]{Dundon05,JH06, KOS07, Licandro07, KJ08, Je13,AMH14}, 
while 1999~YC is a C-type asteroid \citep{KJ08}.  
Heterogeneity on the surfaces of Phaethon and 2005~UD 
may be due to intrinsically  inhomogeneous composition, perhaps affected by hydration processes \citep{Licandro07}, 
and by thermal alteration \citep{KOS07}.    
The rotational color variation of 2005~UD shows B-type for 75\% of 
the rotational phase but C-type for the remainder \citep{KOS07}.  
The colors of the PGC objects are broadly consistent with being neutral-blue.

\index{color}
\index{asteroid!2003~EH$_1$}
\index{comet!96P/Machholz~1}
\index{near-Sun objects} 
Optical colors of 2003~EH$_1$ are taxonomically compatible 
with those of C-type asteroids \citep{KJ15} (Figures~\ref{BVR} and~\ref{VRI}).     
The V-R  color (0.39$\pm$0.01) is similar to  that of
96P \citep[${\it V - R}$ = 0.40$\pm$0.03, from][]{LTL00,MHM04}.  
We note that the optical colors of 2003~EH$_1$
are significantly less red than the average colors of cometary nuclei~\citep{Jewitt02, LTF04}.   
This could be a result of past thermal processing when the object had a perihelion 
far inside Earth's orbit. Indeed, the weighted mean color of 8 near-Sun asteroids having perihelion 
distances $\lesssim$ 0.25 AU (subsolar temperatures $\gtrsim$ 800 K) is V-R = 0.36$\pm$0.01 \citep{Je13}, 
consistent with the color of EH$_1$ (cf. section~\ref{LK}).

\index{color}
\index{comet!169P/NEAT}
\index{Trojan!Jovian}
\index{Bus taxonomy}
\index{taxonomy!T-type}
\index{taxonomy!X-type} 
The optical colors measured for 169P/NEAT are less red than D-type objects, as found in normal 
cometary nuclei and Trojans, 
but similar to those of T- and X- type asteroids (Figures~\ref{BVR} and~\ref{VRI}).  
The near-infrared spectrum measurement (0.8--2.5 $\mu$m) classified 
169P as a T-type asteroid based on the Bus taxonomy with $\it p_v$=0.03$\pm$0.01 \citep{DeMeoBinzel08}.  
The T-type asteroids represent slightly redder-sloped visible wavelength spectra 
than those of C-type.  
Perhaps a refractory rubble mantle has formed on the 169P surface, driven by 
volatile sublimation, and red matter has been lost~\citep{Jewitt02}.

\index{asteroid!1566~Icarus}
\index{color}
\index{taxonomy!Q-type}
\index{taxonomy!V-type} 
Asteroid 1566~Icarus is taxonomically classified as  a Q- or V-type  (Figures~\ref{BVR} and~\ref{VRI}). 
These types suggest thermal evolution (perhaps at the level of the ordinary chondrites) relative to the more primitive carbonaceous chondrites.    
The formation process of the associated complex is unknown, but we speculate that processes other than comet-like sublimation of ice are responsible.

\subsection{Dust Production Mechanism: Example of 1566~Icarus and 2007~MK$_6$}
\label{Rotation}

\begin{figure}
\figurebox{21pc}{}{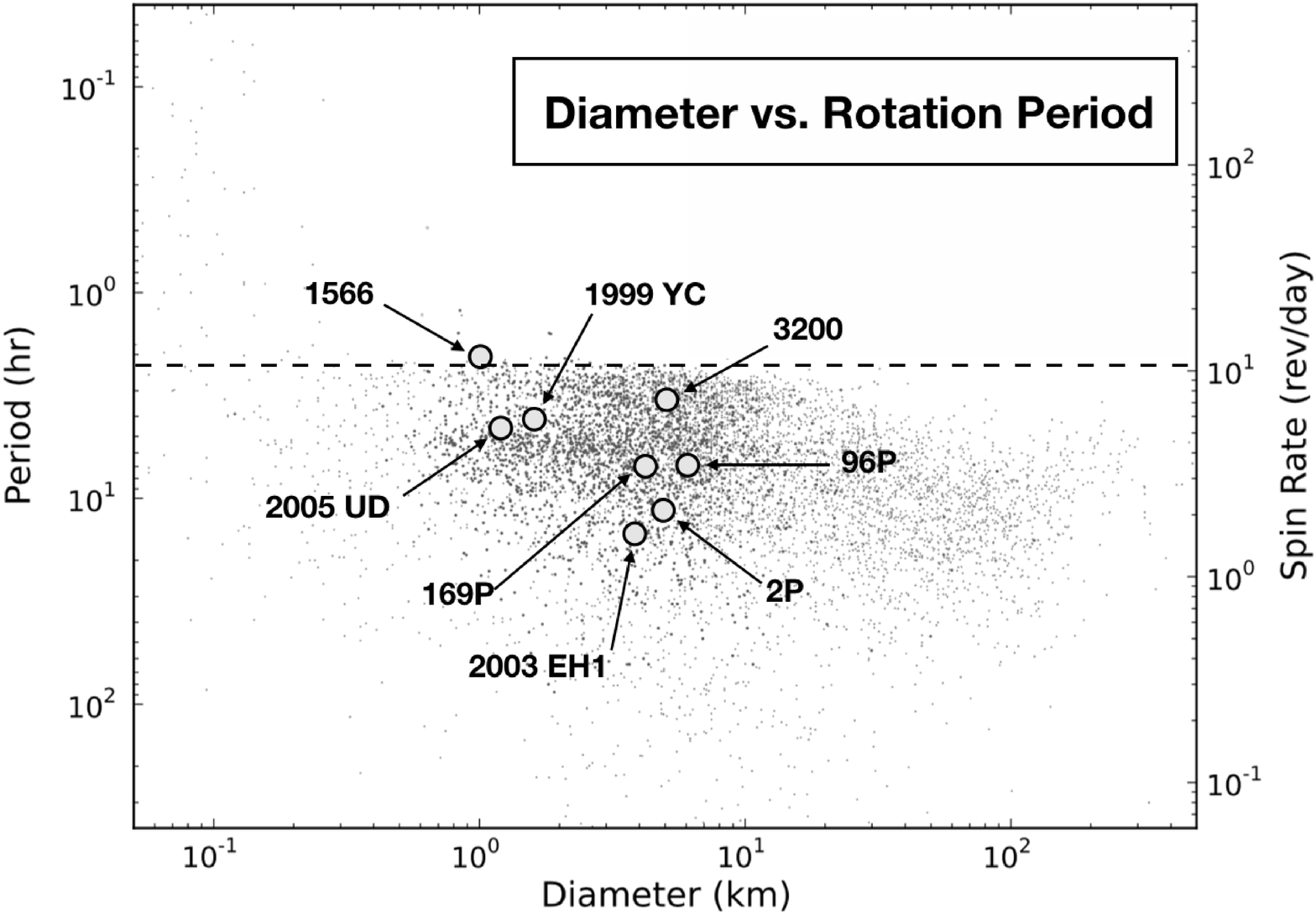}
\caption{Diameter vs.~rotational period for asteroids (dots) and the parent bodies of meteoroid streams (large circles).  The asteroid data are taken from \citet[][]{Chang_Ip_15} while the parent body parameters are listed in Table~\ref{physical}.
The horizontal dotted line shows the spin barrier period of $\sim$2.2\,hr \citep[e.g.][]{WHP09} and asteroid 1566~Icarus rotates nearby.   
} 
\label{DP}
\end{figure}

\index{meteoroid stream!Sekanina's (1973) Taurid-Perseids}
\index{complex!Sekanina's (1973) Taurid-Perseids}
\index{asteroid!1566~Icarus}
\index{asteroid!2007~MK$_6$}
\index{rotational period}
\index{spin-barrier}
Here we consider possible dust production mechanisms from asteroidal parents.  
A diameter -- rotation plot compiled from the data in Table~\ref{physical} is shown in Figure \ref{DP}.
The rotational period of 1566~Icarus ($P_{\rm rot}$=2.273\,hr) is near 
the spin barrier period of $\sim$2.2\,hr \citep{WHP09,Chang_Ip_15}.  
Asteroids rotating near or faster than this barrier are presumed to have been destroyed 
when centrifugal forces have overcome the gravitational and cohesive forces binding them together \citep{Pravec08Icar}.

\index{break-up!rotational instability}
\index{break-up!thermal disintegration}
\index{radiation pressure}
\index{YORP effect}
\index{near-Sun objects}
\index{asteroid!(3200)~Phaethon} 
The aftermaths of  recent and on-going asteroid break-up have been identified observationally \citep[e.g.~P/2013~R3,][]{Jewitt14,Jewitt17} and studied theoretically \citep{Hirabayashi14}. 
Additionally, different mechanisms can operate together.  Rotational instability in P/2013 R3, for instance, might have been induced by YORP torques, or by outgassing torques from sublimated ice, or by a combination of the two.  Thermal disintegration, electrostatic ejection and radiation pressure sweeping may all occur together on near-Sun object 3200~Phaethon \citep[][]{JL10}. 
Figure \ref{DP} shows that 1566 rotates near the $\sim$2.2\,hr spin barrier period, implying a rotational breakup in the past.     
Both 1566 and its possible fragment 2007~MK$_6$ have  small perihelia ($q\sim$0.19\,AU).  We consider rotational instability as a possible cause of their past separation.

%
\index{YORP effect}
\index{near-Earth objects (NEOs)}
\index{collisional lifetime}
In principle, rotation rates of asteroids can be accelerated to critical limits  
by torques exerted from solar radiation through the YORP effect \citep[][]{Vokrouhlick15}.
The YORP e-folding timescale of the spin, $\tau_{\rm Y}$, 
is estimated from the ratio of the rotational angular momentum, $L$, to the torque, $T$.  
The relation may be simply expressed as $\tau_{\rm Y} \sim K$\,$D_e^2$\,$R_h^2$\,\citep{JHA15}, 
where $K$ is a constant, $D_e$ is the asteroid diameter (km) and $R_h$ is the 
heliocentric distance (AU).  
The value of constant $K$ is sensitive to many unknown parameters (the body shape, 
surface texture, thermal properties and spin vector of the asteroid and so on), 
but can be experimentally estimated from published measurements of YORP acceleration 
in seven well-characterized asteroids \citep[Table~2 from][]{RozitisGreen13}.  
Scaling $K$ to the bulk density of 1566~Icarus $\rho$ = 3400\,kg\,m$^{-3}$ 
\citep[V-type or ordinary chondrite from][]{Wilkison00,Britt02} and its rotation period $P_{\rm rot}$=2.273\,hr, 
we find $K \sim$ 7$\times$10$^{13}$\,s\,km$^{-2}$\,AU$^{-2}$.  
The approximation is represented as \citep[cf. Equation (3) of][]{JHA15}, 
\begin{equation}
\tau_Y~(\rm Myr) \approx 2\,\left[\frac{D_e}{\rm 1\,km}\right]^2  \left[\frac{R_h}{\rm 1\,AU}\right]^2.  
\label{yorp}
\end{equation}
For 1566 with $D_e$=1\,km orbiting at $R_h \sim$1.08\,AU,  Equation~($\ref{yorp}$) gives 
$\tau_{\rm Y} \approx$ 2\,Myr.  This is two orders of magnitude smaller than the 
collisional lifetime of 1-km near-Earth asteroids \citep{Bottke94hdtc}, 
suggesting that YORP torque spin-up is plausible.

\index{asteroid!1566~Icarus}
\index{asteroid!2007~MK$_6$}
\index{cohesive strength} 
Asteroids rotating faster than the spin-barrier cannot be 
held together by self-gravitation only, but require cohesive strength \citep[e.g.][]{Scheeres10}.   
The cohesive strength at rotational breakup of a body 
can be estimated by the dispersed fragmental sizes, 
initial separation speed, and the bulk density using Equation (5) of \cite{JHA15}, 
\begin{equation}
S \sim \rho \left(\frac{D_{e}^{'}}{D_{e}}\right)(\Delta v)^2.
\label{cohesive}
\end{equation}
Both fragmental asteroids (1566 and 2007~MK$_6$) are assumed to have the same bulk density $\rho$, 
$D_{e}$ and $D_{e}^{'}$ are diameters of 1566 and 2007~MK$_6$ respectively (see Table~\ref{physical}), 
~and $\Delta v$ is the excess velocity of escaping fragments, assumed comparable to the escape velocity from 1566. 
Adopting the same value for $\rho$ (see above) and substituting 
($D_{e}^{'}$/$D_{e}$) = 0.18 (the diameter ratio between MK$_6$ and 1566), and $\Delta v$ = 0.69\,m\,s$^{-1}$, 
we find $S \sim$ 290\,N\,m$^{-2}$.

\index{asteroid!rubble-pile} 
This small value is comparable to strengths $\sim$10--100\,N\,m$^{-2}$ 
modeled by a rubble-pile asteroid bounded by weak van~der~Waals forces \citep[reviewed in][]{Scheeres18},   
but more than five orders of magnitude smaller than the values of typical competent rocks (10$^7$ -- 10$^8$ N\,m$^{-2}$).  A rotational break-up origin of 1566 and MK$_6$ is possible provided they have a weak, rubble-pile structure, as is thought likely for a majority of kilometer-sized asteroids as a result of past, non-destructive impacts.

%
\index{break-up!thermal disintegration} 
Several processes could eject dust  from the surface of 1566.  Firstly, thermal disintegration can be induced by thermal expansion forces  
that make cracks on the surfaces of asteroids and produce dust particles.  
The characteristic speeds of dust particles produced by disintegration 
can be derived by conversion from thermal strain energy into 
kinetic energy of ejected dust particles.  
The necessary conversion efficiency, $\eta$, is given by \cite[cf. Equation (3) of][]{JL10}
\begin{equation}
\eta \sim \left( \frac{v_e}{\alpha\,\delta T} \right)^2 \left(\frac{\rho}{Y}\right), 
\label{eta}
\end{equation}
where, $v_e$ = 0.69\,m\,s$^{-1}$ is the escape velocity from 1566, 
$\alpha$ $\sim$ 10$^{-5}$\,K$^{-1}$ is the characteristic thermal expansivity of  rock \citep{Lauriello1974,Richter1974}, 
$\delta T \sim$450\,K is the temperature variation between the $q$ and $Q$,  
and $Y$ = (1--10) $\times$ 10$^{10}$ N\,m$^{-2}$ are typical Young's moduli for rock \citep[][p.474]{Pariseau2006}.  
With $\rho$ as above we find $\eta \gtrsim$ 0.1--1\,\% is needed for the velocities of ejected dust particles to surpass the escape velocity.
This very small value of conversion efficiency is  sufficient  
for most dust particles produced by thermal disintegration to be launched into interplanetary space.  

\index{photoionization} 
Secondly, electrostatic forces caused by photoionization by solar UV can eject small particles.  The  critical size for a 1\,km-diameter asteroid  is
 $a_e \lesssim$ 4 $\micron$ \citep[Equation (12)][]{JHA15}.  
Millimeter-sized particles cannot be electrostatically launched and this process may contribute little or nothing to meteoroid stream formation.

\index{radiation pressure} 
Finally, radiation pressure sweeping can remove small particles from an asteroid once they are detached from the surface by another process (i.e.~once the surface contact forces are temporarily broken).   
Radiation pressure sweeping is most effective at small heliocentric distances.  
The critical size to be swept away, $a_{\rm rad}$ ($\micron$), is estimated by equating 
the net surface acceleration (gravitational and centripetal) with the acceleration due to radiation pressure, 
given by Equation (6) of \cite{JL10}
\begin{equation}
a_{\rm rad} \sim \frac{3\,g_\odot}{2\,\pi\,R_{\rm AU}^2\,f^{1/2}\,D_e} \left[\frac{G\,\rho}{f^2} - \frac{3\,\pi}{P_{\rm rot}^2} \right]^{-1}, 
\label{rad}
\end{equation}
where, $g_\odot$ is the gravitational acceleration to the Sun at 1\,AU, 
$R_{\rm AU}$ is the heliocentric distance expressed in AU, 
$f$ is the limit to the axis ratio (=a/b), $G$ is the gravitational constant.  
We substitute $g_\odot$ = 0.006\,m\,s$^{-2}$, 
$R_{\rm AU}$ = 0.187 (Table~\ref{orbit}),
$f$=1.2 (Table~\ref{physical}),
$G$ = 6.67 $\times$ 10$^{-11}$m$^3$\,kg$^{-1}$\,s$^{-2}$ 
and adopt the same values of $D_e$, $\rho$ and $P_{\rm rot}$ (see above) into Equation~(\ref{rad}), 
then obtain $a_{\rm rad}$ $\sim$ 4,500 $\micron$ $\approx$ 5\,mm.  
This size is large enough to contribute meteoroid-sized particles to a stream (see \ref{MS}, Meteoroid Streams).

In summary, asteroid 1566~Icarus is a possible product of  rotational breakup and, given its small perihelion distance, potentially experiences a mass loss process similar to those inferred on Phaethon.   Near-perihelion observations of 1566 and/or 2007~MK$_6$ may indeed show Phaethon-like mass-loss.

\subsection{End State}

\index{dynamical lifetime}
\index{short-period comets} 
Active objects on comet-like orbits with $T_J < $\,3.08 (Table~\ref{orbit} and Figure~\ref{ae}) 
are presumed to be potential ice sublimators.  
The timescales for the loss of ice from a mantled body $\tau_{\rm dv}$,   
for the heat propagation into the interior of a body $\tau_{\rm c}$, 
and dynamical lifetime of short-period comets $\tau_{\rm sp} \sim 10^{5}$--$10^{6}$\,yr \citep[][]{Duncan04} 
can be compared to predict an object's end state \citep[][]{Jewitt04}.

The $\tau_{\rm dv}$ is calculated using $\rho_{\rm n}$$D_e$/2$f$$(dm/dt)$, where 
$\rho_{\rm n}$ = 600\,kg\,m$^{-3}$ is the cometary bulk density \citep{WAL04} and $f$  = 0.01 is the mantle fraction \citep{A'Hearn95}.  
The orbit with averaged $\bar{a} \sim$\,2.7\,AU and $\bar{e} \sim$\,0.8 from the seven objects 
has a specific mass loss rate of water ice $dm/dt \lesssim $\,10$^{-4}$\,kg\,m$^{-2}$\,s$^{-1}$ \cite[Figure~6, in][]{Jewitt04}.
Then we find $\tau_{\rm dv}$~$\gtrsim$~10$^4$\,$D_e$ in yr, where $D_e$ is an effective diameter in km (Table~\ref{physical}).  
The $\tau_{\rm c}$~$\sim$~8.0$\times$10$^4$\,$D_e^2$ in yr is derived from the equation of heat conduction given by  
$D_e^2$/4$\kappa$, where $\kappa$ = 10$^{-7}$\,m$^2$\,s$^{-1}$ is the assumed thermal diffusivity of a porous object. 
The critical size of object to form an inactive, devolatilized surface is constrained by 
the relation $\tau_{\rm dv} \lesssim $ $\tau_{\rm c}$, which gives $D_e \gtrsim$\,0.13\,km. 
Likewise, the size capable of containing ice in the interior of a body for the dynamical lifetime 
is given by the relation $\tau_{\rm c} \gtrsim \tau_{\rm sp}$, which gives $D_e \gtrsim$\,1.1\,km.

\index{asteroid!2003~WY$_{25}$}
\index{comet!D/1819 W1 (289P/Blanpain)}
\index{dormant comet} 
Most comet-like objects are expected to be dormant, with  
ice depleted from the surface region, but potentially still packed deep inside.  
We note that 2003~WY$_{25}$ of the Phoenicids is on its way to the dead state due to its small size ($D_e$ =  0.32\,km).  
The $\tau_{\rm c} \sim$ 8,000\,yr suggests that solar heat can reach into the body core 
approximately 1 or 2  orders of magnitude sooner than the end of the dynamical lifetime.
The core temperature around the orbit \citep{JH06}, $T_{\rm core}$ $\sim$ 180\,K, exceeds 
the sublimation temperature of water ice 150\,K \citep{Yamamoto1985}.   
The extremely weak activity of 2003~WY$_{25}$ may portend its imminent demise (cf. section~\ref{PhoWY25}). 

\begin{table*}
\caption{Closest Approach to the Sun by the Lidov-Kozai Mechanism \label{LKtable}}{\tabcolsep5.5pt%
\begin{tabular}{@{}llccccc @{}}
\toprule 
Complex & Object  &  $e_{\rm max}$\tablenotemark{a} &  $q_{\rm min}\tablenotemark{b}$ & $T_{\rm peak}$\tablenotemark{c} & $ {\rm Ref.}\tablenotemark{d}$\\
\hline                                   
Geminids    & Phaethon        & 0.90 & 0.13 & 1100 & $\sim$0.13$^1$\\                                
            & 2005 UD         &	0.90 & 0.13 & 1100 & 0.13--0.14$^1$\\
            & 1999 YC         &	0.89 & 0.16 & 1000 & - \\  
Quadrantids & 2003 EH$_1$     &	0.96 & 0.12 & 1100 & $\sim$0.12$^2$\\  
            & 96P/Machholz 1  &	0.99 & 0.03 & 2300 & 0.03--0.05$^3$\\   
Taurids-Perseids & 1566 Icarus&	0.85 & 0.16 & 1000 & $\sim$0.17$^4$\\  
                 & 2007 MK$_6$&	0.85 & 0.16 & 1000 & $\sim$0.17$^4$\\  
\hline    
\end{tabular}}
\begin{tabnote} 
Notes: Near-Sun objects have perihelia $\lesssim 0.25$\,AU.\\
 $^a$ Maximum eccentricity (Equation~(\ref{emax}))\\
 $^b$ Minimum perihelion distance (AU) estimated by $\simeq\,a(1-e_{\rm max})$, where $a$ is from Table~\ref{orbit}.\\
 $^c$ Peak temperature at $q_{\rm min}$ (K)\\
 $^d$ Referred minimum perihelion distance (AU) from numerical integrations and analytical methods \\
 $^1$\cite{Ohtsuka97EMP,OSK06} (cf. Figure~\ref{DyUD}) 
 $^2$\cite{NKT13,FG14} (see section~\ref{QUAD}) 
 $^3$\cite{Bailey1992,SC05,AWJ18}  
 $^4$\cite{OAI07} (cf. Figure~\ref{MK6}) 
\end{tabnote}  
\end{table*}

\subsection[Lidov-Kozai Mechanism]{Lidov-Kozai Mechanism\protect
\footnote{In Chapter~7 \citep{chapter7} the Lidov-Kozai mechanism focuses on secular changes in $e$, $i$ and $\omega$ 
to find how $e$ and $\omega$ relate to whether an orbit intersects Earth's orbit to produce a meteor shower.}}
\label{LK}

\index{Lidov-Kozai mechanism}
The Lidov-Kozai mechanism works on the secular dynamics of small solar system objects. 
Large-amplitude periodic oscillations of the $e$ and $i$ (in antiphase) are produced,  
whereas the $a$ is approximately conserved, 
while the $\omega$ librates around $\pi/2$ or $3\pi/2$ if $C_2 < 0 $ 
or circulates if $C_2 > 0$ 
\citep{Kozai62,Lidov62}. 

\index{Lidov-Kozai mechanism}
\index{thermal process}
\index{near-Sun objects} 
Perihelia can be deflected into the vicinity of the Sun 
by this mechanism (on timescale $\sim$1000s of years), perhaps causing physical alteration (or even breakup) 
due to enormous solar heating  \citep{Emel'yanenko17}. 
We find all parent bodies stay in the circulation region ($C_2 > 0$, see Table~\ref{orbit}).
The minimum perihelion distance $q_{\rm min}$ can be computed using 
maximum eccentricity, $e_{\rm max}$, given by (Equations (5) and (28) of \cite{Antognini2015})
\begin{equation}
e_{\rm max} = \sqrt{1-\frac{1}{6} \left(\zeta - \sqrt{\zeta^2 - 60\,C_1} \right)}, 
\label{emax}
\end{equation}
where $\zeta = 3 + 5(C_1 + C_2)$ \cite[Equation (31) of][]{Antognini2015}, 
$C_1$ and $C_2$ are from Table~\ref{orbit}.  
With the obtained $q_{\rm min}$, we find 
the Geminids (PGC), Quadrantids and Sekanina's (1973) Taurid-Perseids complexes 
are near-Sun objects (Table~\ref{LKtable}).
Amongst them, 2003~EH$_1$ turns itself into a near-Sun object
with $q_{\rm min}$$\sim$ 0.12\,AU (cf. section~\ref{QUAD}), 
albeit $q$ $\sim$ 1.2\,AU at present (see Figure~\ref{ae}).  
The peak temperature~$\gtrsim$~1000\,K is similar to that experienced by Phaethon (cf. 1566~Icarus),  
and could likewise cause strong thermal and desiccation stresses, cracking and alteration in EH$_1$, with the release of dust  \citep[][]{JL10,Molaro15,Springmann18}.  
This example reminds us that, even in objects with $q$ presently far from the Sun, we cannot exclude the action of extreme thermal processes in past orbits.


\section{Meteors and Streams}

\subsection{Na Loss: Thermal}
\label{Na}

\begin{figure}
\figurebox{20pc}{}{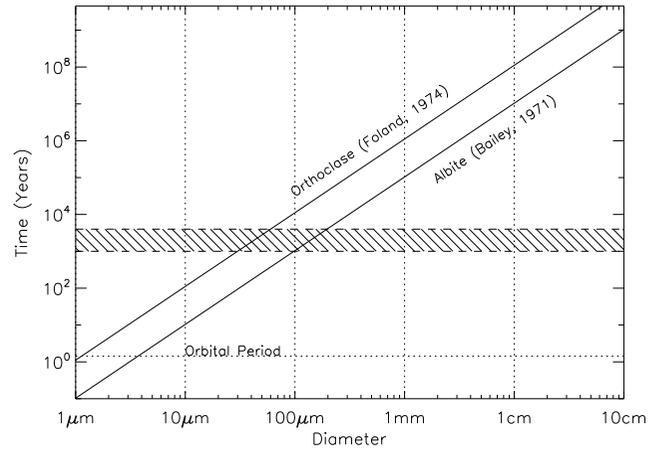}
\caption{
Timescale for escape of 90\% of initial Na content from Geminid meteoroids in the stream, 
as a function of the meteoroid size.    
The low diffusion data for Na in orthoclase was used as a limit for the slowest loss, 
while the faster diffusion data for Na in albite (10$\times$ higher than for orthoclase) 
is considered as a more realistic value for the Geminids.
The time interval in the shaded region corresponds to the estimated age of the Geminid meteoroid stream (1000--4000\,yr).   
From \cite{DB09}.
\label{Capek2009}}
\end{figure}

\index{sodium (Na)}
\index{abundance!solar} 
Sodium loss in Geminid meteors results from the action of a thermal process in or on the parent body, Phaethon
(section~\ref{PGC}). 
\cite{DB09} calculated the timescale for thermal depletion of Na 
from an assumed initial solar value down to 10\% of solar abundance in Geminid meteoroids 
 during their orbital motion in interplanetary space (using assumed albite ($\rm NaAlSi{_3}O{_8}$) and orthoclase ($\rm KAlSi{_3}O{_8}$) compositions).
With particle diameters~$\ge$~mm-scale, they found depletion timescales~$\gtrsim$~10$^{4}$--10$^{5}$\,yr  (Figure~\ref{Capek2009}), some  
1 to 2 orders of magnitude longer than the stream age of $\lesssim$~10$^{3}$ yr.  
On the other hand, the dynamical lifetime of Phaethon, while very uncertain, is estimated to be $\sim$ 3 $\times$ 10$^7$\,yr \citep{deLeon10}. 
This is surely long compared to the age of the Geminid stream and 
long enough for Na to be thermally depleted from Phaethon.

\index{sputtering}
\index{solar wind}
\index{planet!Mercury} 
In principle, other processes might affect the sodium abundance.  Sputtering by the solar wind, photon stimulated Na desorption (on 
Mercury and the Moon) \citep{McGrath86,PKM00,KSP04,YM04} and cosmic ray bombardment \citep[]{SNH01} have been well studied.  
These processes act only on the surface, and are inefficient in removing Na from deeper layers \citep{DB09}.

\index{abundance}
\subsection{Abundances vs. Intensity Ratios}
\label{AvsI}

\index{abundance!elemental}
\index{intensity ratio}
\index{abundance!solar}
\index{main component}
\index{hot component}
\index{Saha function} 
In the interpretation of meteor spectra, 
two types of evaluation methods are employed in the literature.   
Some investigators calculate elemental abundances while others use simple line intensity ratios.  
Here we describe advantages and disadvantages hidden  
in both methods. 

Elemental abundances are quantitatively utilized for comparing with 
those of meteor showers and the solar abundances \citep{AG89,L03}. 
The derivation of abundances is influenced by the complicated physics of the ionized gas at the head of the meteor \citep{Bo93, KYW05}.  
For example, the Saha equation is needed to calculate the neutral vs.~ion balance,  but this depends on the assumption of a
 plausible excitation temperature, $T_{ex}$, for the emitting region. 
This is particularly important for Na, which has a smaller first ionization energy ($\approx$ 5.1\,eV) compared to other species (e.g.~Mg: $\approx$ 7.6\,eV, Fe: $\approx$ 7.9\,eV). The selection of the appropriate $T_{ex}$ is problematic.  
Fireball-like spectra have been suggested to be the combination of some thermal components with typical $T_{ex}$ $\sim$ 5,000, 8,000 and 10,000\,K, respectively \citep{Bo93,KYW05, KIW07}. 
\cite{Bo93} considered Ca\,{\sc ii} lines in the hot components 
($T_{ex}$ $\sim$ 10,000\,K) and estimated electron density from 
the radiating volume of the meteor in the direction of its flight.   
This implies that Ca\,{\sc ii} may not reflect the actual electron density from their spectral emission profile.  
In \cite{KYW05}, on the other hand, Ca\,{\sc ii} is taken from the main component 
($T_{ex}$ $\sim$ 5,000\,K) instead of the hot one, and they derive electron density 
reflected upon the measured spectral profile.

The definition of hot component theory has to 
satisfy the equality of total metal abundances (Ca/Mg) and 
pressure of the radiant gas between the main and the hot components \citep{Bo93}. 
However, the relation does not fit in most spectral data (e.g. Leonids).  
The Saha function is used to verify the definition but mostly finds negative values 
of electron density, which is clearly unrealistic \citep{KYW05}. 
This proves that the original hot component theory may go against its own definition \citep{Bo93}.  
\cite{KYW05} suggest that the Ca\,{\sc ii} lines do not always belong to the hot component, 
but instead to the main component.  
Plausibility is also found in their low excitation energies (Ca\,{\sc ii} $\approx$ 3.1\,eV), 
which are compatible with those of other neutral metals (e.g. Na\,{\sc i} $\approx$ 2.1\,eV) 
identified in the main component \citep{KWK07}.   
On the other hand, the hot component primarily consists of species with high excitation energies $\gtrsim$ 10\,eV.  
Accordingly the Ca\,{\sc ii} lines are most likely to belong to the main component.

Given the difficulties in calculating absolute abundances, many researchers have employed simple line intensity ratios \citep[e.g.][]{BKS05}.    
The method analyzes neutral atomic emission lines of Na\,{\sc i}, Mg\,{\sc i}, and (weak) Fe\,{\sc i} only.
Note that the intensity ratios do not directly reflect the elemental abundances 
due to no consideration for excitation properties of the elements and ions, including electron densities \citep[e.g.][]{B01,BKS05,KBS06}.  
Laboratory spectroscopic experiment proves that intensity ratios are not informative of the abundance 
and cannot be used to determine the meteorite analogue \citep{Drouard18}.
Line ratios can be used, at best, to study the  trend of elemental content in meteors which happen to possess similar physical conditions (including similar entry velocities, similar strengths and similar excitation temperatures).  Line intensity ratios can suggest  the trend of elemental content in meteor showers but cannot provide  the abundances.

\index{meteoroid stream}
\subsection{Meteoroid Streams}
\label{MS}

\index{meteoroid stream}
\index{ablation}
Physical properties of meteoroids in meteoroid streams or 
debris in dust trails whose orbits do not intersect that of the Earth 
can be revealed by thermal and optical observations.   
The streams (or trails) and meteoroids mostly consist of mm to cm-scale compact aggregates, 
as estimated from the ratio of solar radiation pressure to solar gravity 
of $10^{-5}$~$\lesssim$~$\beta$~$\lesssim$~$10^{-3}$ 
\citep[e.g.][]{SykesWalker92,RSL2000,Reach07,Ishiguro02,Sarugaku15}.

Meteor ablation models in the Earth atmosphere are also 
available to estimate the size of meteoroids \citep[][]{Bron83,CBE98}.  
The classical models find typical meteors are 10$\mu$m -- 10\,cm in size, 
however, with uncertainties caused by various parameters 
(e.g. luminous efficiency, ablation coefficient, and fragmentation) 
which are sensitive to the meteoroid entry speed, tensile strength and brightness  
\citep[][]{CBE98,Baba2002AA}.  
Faint meteors are estimated to be 10$\mu$m $\sim$1\,mm, 
but model improvements are needed to better represent fragmentation 
\citep{CampbellKoschny04,Borovi07}.

\subsection{Zodiacal Cloud} 

\index{zodiacal cloud}
The zodiacal cloud is a circumsolar disk consisting of small dust particles supplied by comets and asteroids.  
The total mass is $\sim$ 4 $\times$ 10$^{16}$\,kg, most  ($\sim$90\%?) of which is  supplied by JFC disruptions 
and the rest  by Oort cloud comets ($\lesssim$ 10\%) 
and asteroids ($\lesssim$ 5\%) \citep{Nesvorny10,Nesvorny11Zodiacal,Jenniskens15aste.book}.  The supply rate needed to maintain the zodiacal cloud in steady-state is 10$^3$ to 10$^4$ kg s$^{-1}$ \citep{Nesvorny11Zodiacal}.

\index{Jupiter family comets (JFCs)}
\index{collisional lifetime}
\index{meteors!sporadic}
\index{Poynting-Robertson drag}
\index{solar wind}
\index{Oort cloud comets}
\index{meteoroids!$\beta$-meteoroids} 
The fate of dust particles released from comets into the zodiacal cloud is traceable \citep[cf.][]{Grun85}.
Sub-micron particles, with $\beta$ $>$ 0.5, are immediately blown out of the solar system on hyperbolic orbits 
by radiation pressure and are referred-to as $\beta$-meteoroids \citep{ZookBerg75,Grun85}.
The JFCs frequently disintegrate when near perihelion
and form dust trails or meteoroid streams, mainly consisting of mm -- cm sized dust particles (see section~\ref{MS}).  
The collisional lifetime of mm-scale particles at 1\,AU is estimated to be $\tau_{\rm col}$~$\gtrsim$~10$^{5}$\,yr, 
modeled with the orbital distribution of sporadic meteors measured by radar 
\citep{Nesvorny11Zodiacal} \citep[cf. $\tau_{\rm col}$ $\sim$ 10$^{4}$ or 10$^{5}$ yr,][]{Grun85,Soja16,YangIshi18}. 
On the other hand, Poynting-Robertson (P-R) and solar-wind drag cause dust particles to spiral down to the Sun.  
The P-R drag timescale, $\tau_{\rm PR}$, to drift down from $\bar{a} \sim$ 2.7\,AU with $\bar{e} \sim$ 0.8 (objects with $T_J < $\,3.08, see Table~\ref{orbit} and Figure~\ref{ae})
to 1\,AU around the Earth orbit ($e \sim$ 0.017) is calculated using the equation in \cite{WyattWhipple1950} \citep[cf.][]{Dermott02}, 

\begin{equation}
\tau_{\rm PR} \simeq \frac{730}{\beta(1 + sw)}~{\rm yr}, 
\end{equation} 

\noindent where $\beta$ $\lesssim$ 10$^{-3}$ (see~\ref{MS}) for dust particles of radius $\gtrsim$ 1\,mm,  with bulk density of 600\,kg\,m$^{-3}$ \citep{WAL04}, and
$sw$ = 0.3 is efficiency of solar-wind drag on a particle normalized to the P-R drag effect \citep{Gustafson94}.  
The estimated P-R drag lifetime is $\tau_{\rm PR}$ $\gtrsim$ 6$\times$10$^5$ yr.
This being somewhat longer than $\tau_{\rm col}$, the mm-scale dust particles are  subject to  collisional disruption while spiralling  down to the Sun by P-R drag.  
As a result of competition between these two effects (loss to Poynting-Robertson at small sizes, loss to collisional shattering at large sizes), the 100--200$\mu$m dust particles are the most abundant in the zodiacal cloud \citep{Love550,Grun85,CBE98,Nesvorny10}.

\index{collisional lifetime}
\index{packing effect}
\cite{Nesvorny11Zodiacal} noted that the collisional lifetime for mm-scale particles 
is long compared to the plausible lifetimes of most meteoroid streams ($\lesssim$ 10$^4$\,yr).
They speculate that cm-scale particles are sources of smaller dust grains.  
Centimeter-scale particles are also released from JFCs.  
The sequence may result in a population of mm-sized or smaller particles which could be more resistant to collisions.  
Recent meteor observations suggest a relative lack of large particles ($\sim$7\,mm) \citep{Jenniskens2016c},
and also suggest that some of these larger particles disappear on timescales $\sim$10$^4$\,yr,
not from collisions, but from other processes.
\cite{Moorhead17} finds a two-population sporadic meteoroid bulk density distribution 
 suggesting that the physical character of freshly ejected dust particles could be altered over time. 
As another example, Rosetta dust collectors sampled both very pristine fluffy aggregates and compact particles ($\gtrsim$\,4\,cm in diameter)
with a possible range of the dust bulk density from 400 to 3000\,kg\,m$^{-3}$ \citep{RSD15}.  
This variety could have resulted from aggregate fragmentation into the denser collected grains as the spacecraft approached, 
while the packing effect is proposed as a plausible mechanism theory for fluffy dust particles released from comets \citep{MukaiFechtig83}.  
The particle size could be reduced on a timescale of 10$^4$ --10$^5$\,yr, comparable with the meteoroid stream lifetime.  
The effect makes the bulk density increase from 600\,kg\,m$^{-3}$ to 3000\,kg\,m$^{-3}$, 
corresponding to shrinking the particle size approximately down to half.  
This could be a potential explanation for disappearing larger-scale dust particles in the meteoroid streams. 


\section{Summary and Future Work}
In the last decade, a growing  understanding of parent bodies and meteoroid streams has been achieved by 
combining new physical observations and dynamical investigations.   
Still, even where the associations are relatively clear, most complexes have multiple potential parent bodies and it remains unclear how the streams were formed.

\index{near-Earth objects (NEOs)} 
Observationally, a major challenge is posed by the difficulty of measuring the physical characteristics of parent bodies,  most of which are faint by virtue of their  small size (typically $\lesssim$ a few \,km). They are also frequently observationally inaccessible because of their eccentric orbits, which cause them to spend most of the time far away near aphelion.     
Long-term surveillance of NEOs around their entire orbits  might better reveal how and when parent bodies disintegrate and produce debris.  

\index{D-criterion} 
Dynamically, there are at least two challenging problems.  One concerns the   identification of parent bodies through the comparison of the orbital elements of meteoroids and potential parents by a D-criterion.  Such methods work best for parents of young streams, where the effects of differential dynamical evolution are limited.  
However, in older systems, the dynamical elements have evolved enough to seriously undercut the use of the D-criteria.  
For this reason, for example, numerous Taurid parent bodies continue to be proposed.  A key objective is to find a way to more reliably associate older meteoroid streams
with their parent bodies. 
A second problem is the use of  long-term dynamical simulations in which the initial conditions and/or potentially important non-gravitational effects are partly or wholly neglected.

\index{meteoroid stream!Geminids}
\index{asteroid!(3200)~Phaethon}
\index{asteroid!2005~UD}
\index{asteroid!1999~YC}
\index{color}
\index{complex!Phaethon-Geminid Complex (PGC)}
\index{meteoroid stream!Quadrantids}
\index{asteroid!2003~EH$_1$}
\index{comet!96P/Machholz~1}
\index{meteoroid stream!Capricornids}
\index{asteroid!2017~MB$_1$}
\index{dormant comet}
\index{complex!Taurid Complex}
\index{comet!2P/Encke}
\index{meteoroid stream!Sekanina's (1973) Taurid-Perseids}
\index{asteroid!1566~Icarus}
\index{asteroid!2007~MK$_6$}
\index{meteoroid stream!Phoenicids}
\index{meteoroid stream!Andromedids}
\index{near-Sun objects}
\index{Lidov-Kozai mechanism}
\index{spectroscopy}
\index{meteoroids!sporadic} 

We list key questions to be answered in the next decade.    

\begin{enumerate}

\item  Geminids:  
What process can act on $\sim$1000 yr timescales to produce the Geminid meteoroid stream?  Phaethon appears dynamically associated with at least two kilometer-sized asteroids (2005~UD and 1999~YC) suggesting a past breakup or other catastrophe.  But the likely timescale for such an event is $\gg$1000 yr.  What caused the breakup and is it related to the Geminids?
How many other PGC-related objects await discovery?  Are Geminids represented in the meteorite collections and, if so, how can we identify them?

\item Quadrantids: 
Presumed parent 2003~EH$_1$ is currently inactive, but was recently as close to the Sun as is Phaethon at perihelion. 
Can residual mass loss in EH$_1$ be detected? Is the Quadrantid sodium abundance depleted as a result of the previously smaller perihelion? 
What physical difference is to be found in 96P which has a near-Sun orbit even now?  

\item Capricornids: 
Several parent bodies have been proposed including both active comets and inactive asteroids.  Did they originate from a common precursor? 
Is asteroid 2017~MB$_1$ related?

\item Taurids: 
The prime parent body is 2P/Encke but  numerous additional  parents with diverse properties continue to be proposed (mostly based on the D-criterion).  How can we establish the relevance of these other objects to the Taurid stream?  Can activity be detected? 
Is the D-criterion appropriate to judge?  

\item Sekanina's (1973) Taurid-Perseids: Do 1566~Icarus and 2007~MK$_6$ share common physical properties?  
Is the sodium abundance in Taurid-Perseids depleted by solar heating due to the small perihelion?

\item Phoenicids, Andromedids and other minor complexes:   
Fragmentation is expected to produce a wide range of object sizes, with many bodies being too small to have been detected so far.   
What role can be played in the search for stream-related bodies by upcoming deep sky surveys, like the Large Synoptic Survey Telescope?   

\item Which is the better index, the D-criterion (e.g. D$_{SH}$) or the dynamical invariants ($C_1$, $C_2$)?

\item How many near-Sun objects, driven by the Lidov-Kozai mechanism, exist?

\item What more can we learn from meteor spectroscopy, particularly of faint meteors?

\item Sporadic meteoroid populations tend to lose large dust particles (sizes $\gtrsim$ 7mm) on timescales of 10$^4$\,yr. Why?

\end{enumerate}

\begin{acknowledgment}
\section*{Acknowledgments}
We are grateful to David Asher for his enthusiastic guidance to improve this chapter. 
TK thanks Takaya Okamoto for wholehearted assistance with this study and also Junichi Watanabe, Hideyo Kawakita, Mikiya Sato and Chie Tsuchiya for support.
We appreciate David {\v C}apek, Ji{\v{r}}{\'\i} Borovi{\v{c}}ka, Man-To Hui, Jing Li and Michael S. P. Kelley for figure contributions.     
We acknowledge reviews by Tadeusz Jopek and an anonymous reviewer.   
Lastly, we thank Galina Ryabova, again David Asher and Margaret Campbell-Brown for organizing this Meteoroids-book project.

\end{acknowledgment}
\backmatter
\bibliography{reference}\label{refs}
\bibliographystyle{cambridgeauthordate-six-au}

\end{document}